\documentclass[twoside,twocolumn,9pt]{article}
\usepackage{extsizes}
\usepackage[super,sort&compress,comma]{natbib}
\usepackage[version=3]{mhchem}
\usepackage[left=1.5cm, right=1.5cm, top=1.785cm, bottom=2.0cm]{geometry}
\usepackage{balance}
\usepackage{times,mathptmx}
\usepackage{sectsty}
\usepackage{graphicx} 
\usepackage{lastpage}
\usepackage[format=plain,justification=justified,singlelinecheck=false,font={stretch=1.125,small,sf},labelfont=bf,labelsep=space]{caption}
\usepackage{float}
\usepackage{fancyhdr}
\usepackage{fnpos}
\usepackage[english]{babel}
\addto{\captionsenglish}{%
  \renewcommand{\refname}{Notes and references}
}
\usepackage{array}
\usepackage{droidsans}
\usepackage{charter}
\usepackage[T1]{fontenc}
\usepackage[usenames,dvipsnames]{xcolor}
\usepackage{setspace}
\usepackage[compact]{titlesec}
\usepackage{hyperref}

\usepackage{amsmath}
\usepackage{amssymb}
\usepackage{siunitx}
\usepackage{multirow}

\newcommand\ie{i.e.\ }
\newcommand\eg{e.g.\ }

\usepackage[textsize=small]{todonotes}

\usepackage{epstopdf}

\definecolor{cream}{RGB}{222,217,201}

\begin{document}

\pagestyle{fancy}
\thispagestyle{plain}
\fancypagestyle{plain}{

\fancyhead[C]{\includegraphics[width=18.5cm]{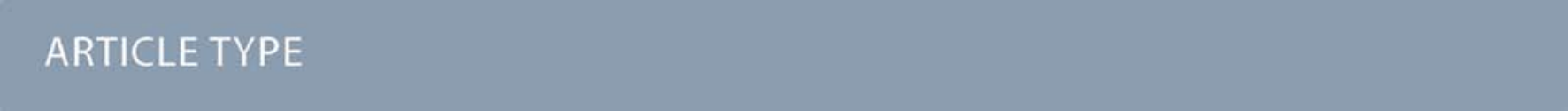}}
\fancyhead[L]{\hspace{0cm}\vspace{1.5cm}\includegraphics[height=30pt]{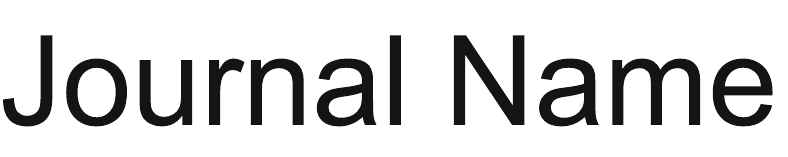}}
\fancyhead[R]{\hspace{0cm}\vspace{1.7cm}\includegraphics[height=55pt]{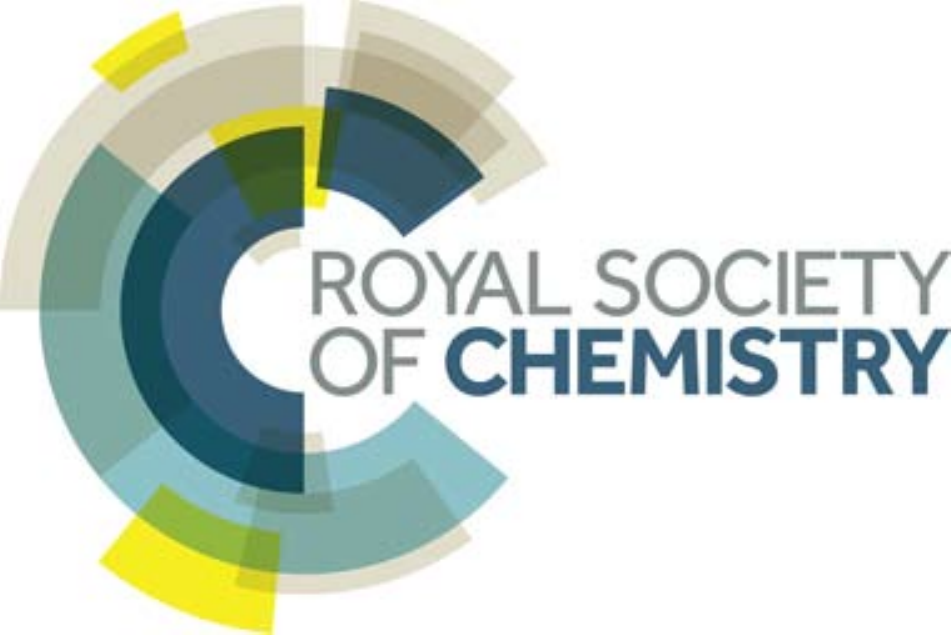}}
\renewcommand{\headrulewidth}{0pt}
}

\makeFNbottom
\makeatletter
\renewcommand\LARGE{\@setfontsize\LARGE{15pt}{17}}
\renewcommand\Large{\@setfontsize\Large{12pt}{14}}
\renewcommand\large{\@setfontsize\large{10pt}{12}}
\renewcommand\footnotesize{\@setfontsize\footnotesize{7pt}{10}}
\makeatother

\renewcommand{\thefootnote}{\fnsymbol{footnote}}
\renewcommand\footnoterule{\vspace*{1pt}%
\color{cream}\hrule width 3.5in height 0.4pt \color{black}\vspace*{5pt}} 
\setcounter{secnumdepth}{5}

\makeatletter 
\renewcommand\@biblabel[1]{#1}            
\renewcommand\@makefntext[1]%
{\noindent\makebox[0pt][r]{\@thefnmark\,}#1}
\makeatother 
\renewcommand{\figurename}{\small{Fig.}~}
\sectionfont{\sffamily\Large}
\subsectionfont{\normalsize}
\subsubsectionfont{\bf}
\setstretch{1.125} 
\setlength{\skip\footins}{0.8cm}
\setlength{\footnotesep}{0.25cm}
\setlength{\jot}{10pt}
\titlespacing*{\section}{0pt}{4pt}{4pt}
\titlespacing*{\subsection}{0pt}{15pt}{1pt}

\fancyfoot{}
\fancyfoot[LO,RE]{\vspace{-7.1pt}\includegraphics[height=9pt]{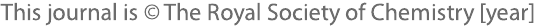}}
\fancyfoot[CO]{\vspace{-7.1pt}\hspace{13.2cm}\includegraphics{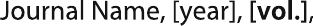}}
\fancyfoot[CE]{\vspace{-7.2pt}\hspace{-14.2cm}\includegraphics{head_foot/RF}}
\fancyfoot[RO]{\footnotesize{\sffamily{1--\pageref{LastPage} ~\textbar  \hspace{2pt}\thepage}}}
\fancyfoot[LE]{\footnotesize{\sffamily{\thepage~\textbar\hspace{3.45cm} 1--\pageref{LastPage}}}}
\fancyhead{}
\renewcommand{\headrulewidth}{0pt} 
\renewcommand{\footrulewidth}{0pt}
\setlength{\arrayrulewidth}{1pt}
\setlength{\columnsep}{6.5mm}
\setlength\bibsep{1pt}

\makeatletter 
\newlength{\figrulesep} 
\setlength{\figrulesep}{0.5\textfloatsep} 

\newcommand{\topfigrule}{\vspace*{-1pt}%
\noindent{\color{cream}\rule[-\figrulesep]{\columnwidth}{1.5pt}} }

\newcommand{\botfigrule}{\vspace*{-2pt}%
\noindent{\color{cream}\rule[\figrulesep]{\columnwidth}{1.5pt}} }

\newcommand{\dblfigrule}{\vspace*{-1pt}%
\noindent{\color{cream}\rule[-\figrulesep]{\textwidth}{1.5pt}} }

\makeatother

\twocolumn[
  \begin{@twocolumnfalse}
\vspace{3cm}
\sffamily
\begin{tabular}{m{4.5cm} p{13.5cm} }
\includegraphics{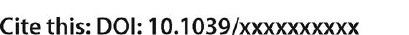} & \noindent\LARGE{Inflow boundary conditions determine T-mixer efficiency\dag} \\
\vspace{0.3cm} & \vspace{0.3cm} \\

 & \noindent\large{Tobias Schikarski,\textit{$^{a}$} Holger Trzenschiok,\textit{$^{a}$} Wolfgang Peukert,\textit{$^{a}$} and Marc Avila\textit{$^{b}$}} \\
\includegraphics{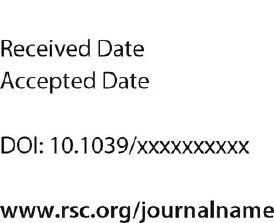} & \noindent\normalsize{We report on a comprehensive experimental-computational study of a simple T-shaped mixer for Reynolds numbers up to $4000$. In the experiments, we determine the mixing time by applying the Villermaux--Dushman characterization to a water-water mixture. In the numerical simulations, we resolve down to the smallest (Kolmogorov) flow scales in space and time. Excellent agreement is obtained between the experimentally measured mixing time and numerically computed intensity of segregation, especially in the turbulent regime, which validates both approaches. We confirm that the mixing time is mainly determined by the specific power input, as assumed in most mixing-models. However, we show that by suitably manipulating the inflow conditions, the power input necessary to achieve a given mixing time can be reduced by a factor of six. Our study enables detailed investigations of the influence of hydrodynamics on chemical reactions and precipitation processes, as well as the detailed testing of turbulence and micromixing models.
} \\

\end{tabular}

 \end{@twocolumnfalse} \vspace{0.6cm}

  ]

\renewcommand*\rmdefault{bch}\normalfont\upshape
\rmfamily
\vspace{-1cm}


\footnotetext{\textit{$^{a}$~Institute of Particle Technology, Department of Chemical and Biological Engineering, Friedrich-Alexander-Universit\"at Erlangen-N\"urnberg, 91058 Erlangen, Germany; E-mail: tobias.schikarski@fau.de}}
\footnotetext{\textit{$^{b}$~Center of Applied Space Technology and Microgravity, Universit\"at Bremen, Am Fallturm, 28359 Bremen, Germany }}

\footnotetext{\dag~Electronic Supplementary Information (ESI) available: [details of any supplementary information available should be included here]. See DOI: 10.1039/b000000x/}



\section{Introduction}
\begin{figure*}
\begin{center}
\begin{tabular}{cccccc}
\multicolumn{3}{c}{(a)} & \multicolumn{3}{c}{(b)}  \\
\multicolumn{3}{c}{\includegraphics[width=0.5\textwidth]{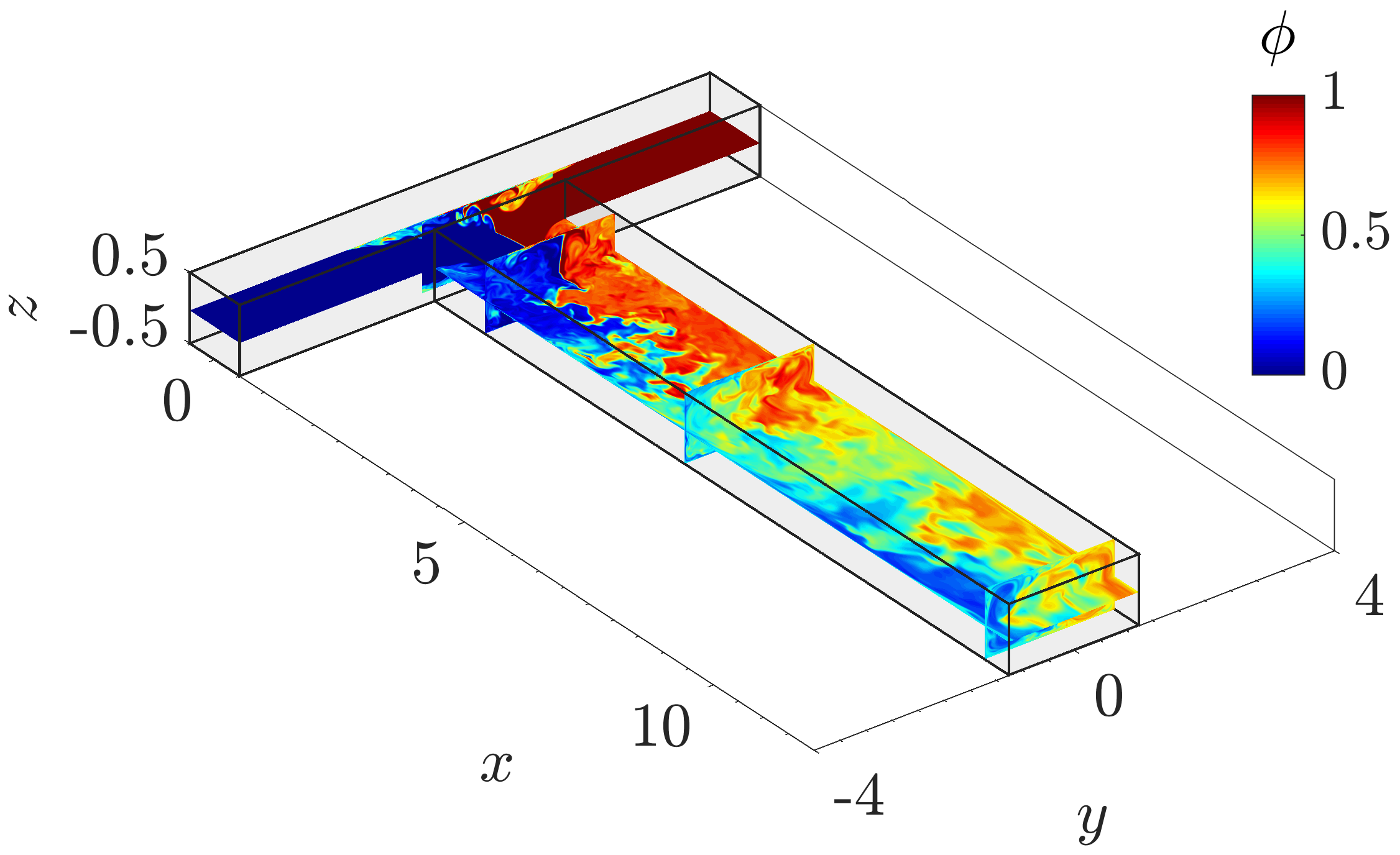}} &
\multicolumn{3}{c}{\includegraphics[width=0.40\textwidth]{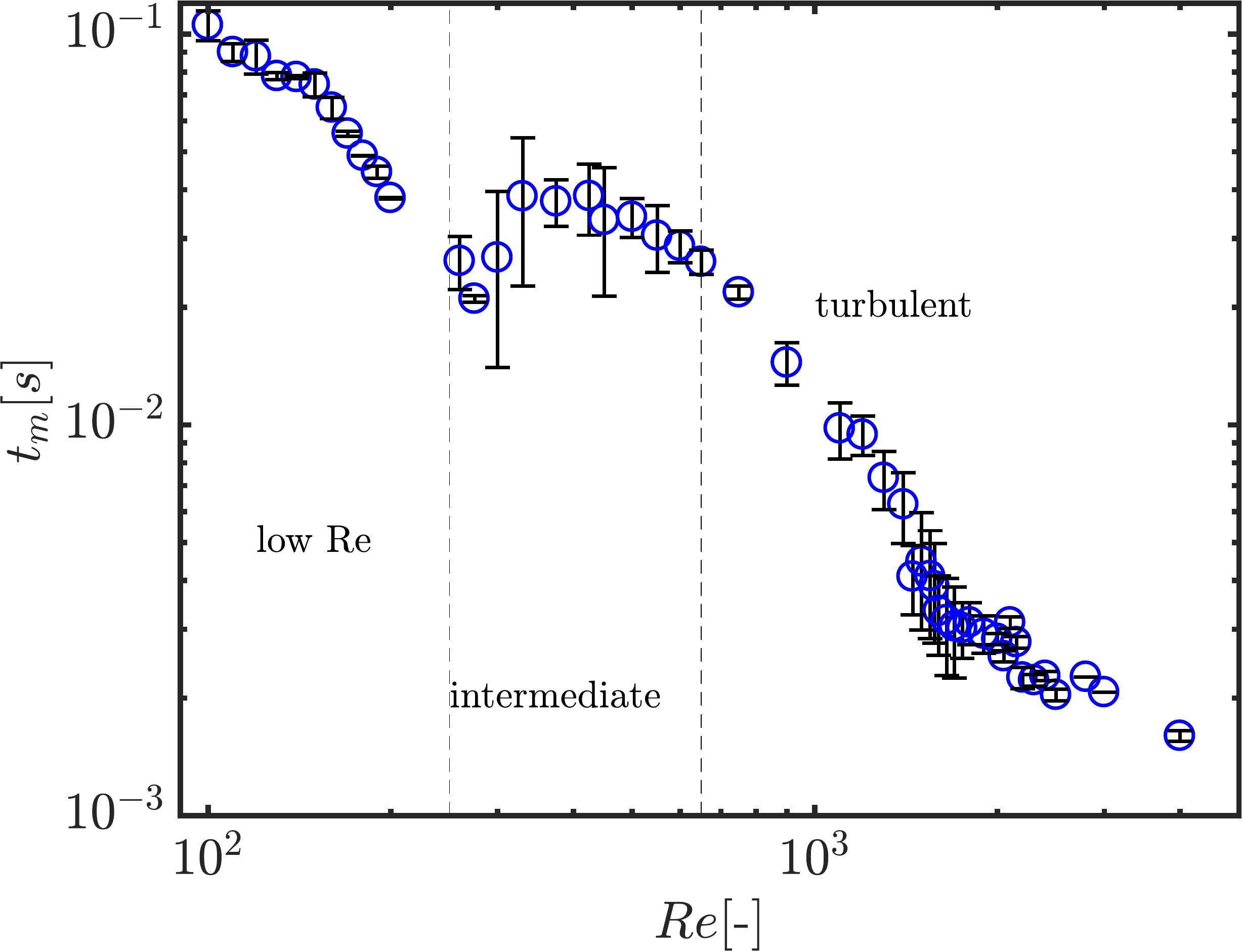}}\\
(c) &  (d) & (e) & (f) & (g) & (h)\\
vortex & engulfment & engulfment & symmetric & turbulent & turbulent \\
$Re=90$ & $140$ & $330$, $t\approx670d/u_0$ & $330,$  $t\approx860d/u_0$ & $900$ & $4000$\\
\includegraphics[width=0.12\textwidth]{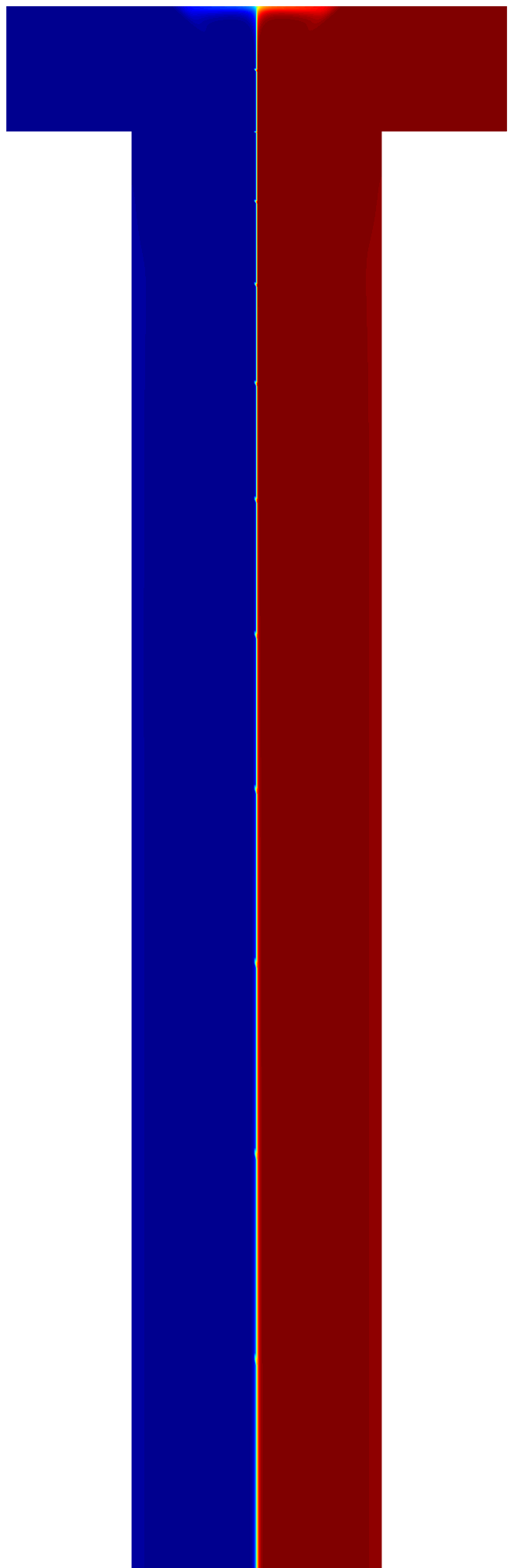} &
\includegraphics[width=0.12\textwidth]{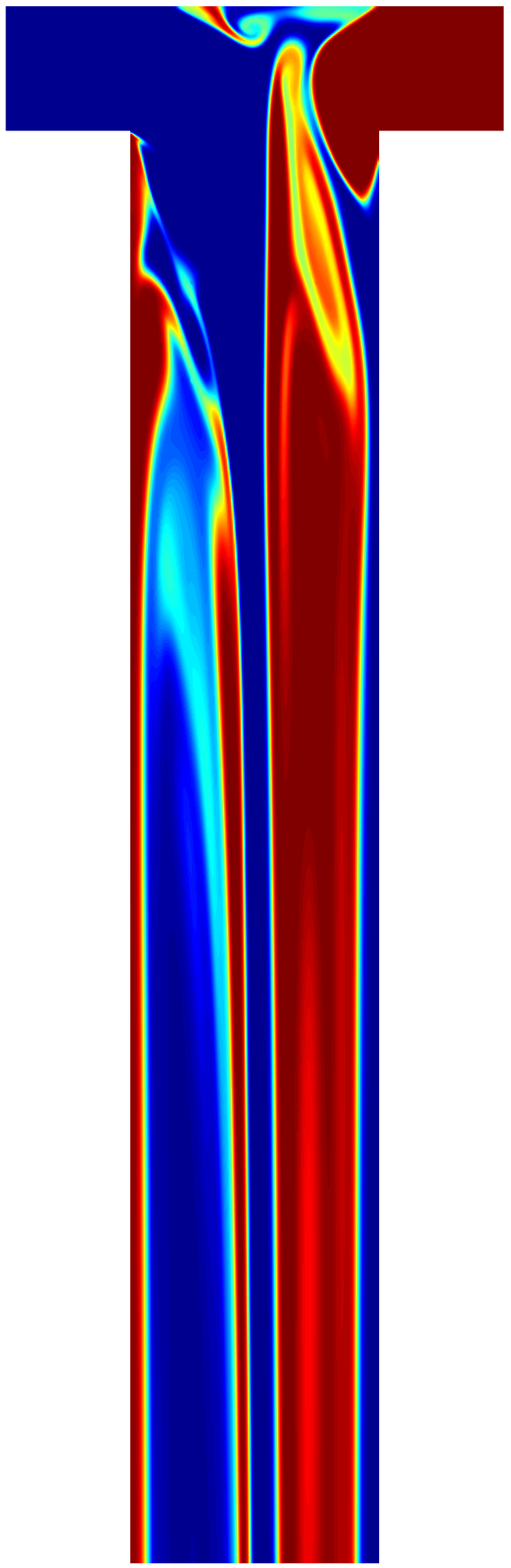} &
\includegraphics[width=0.12\textwidth]{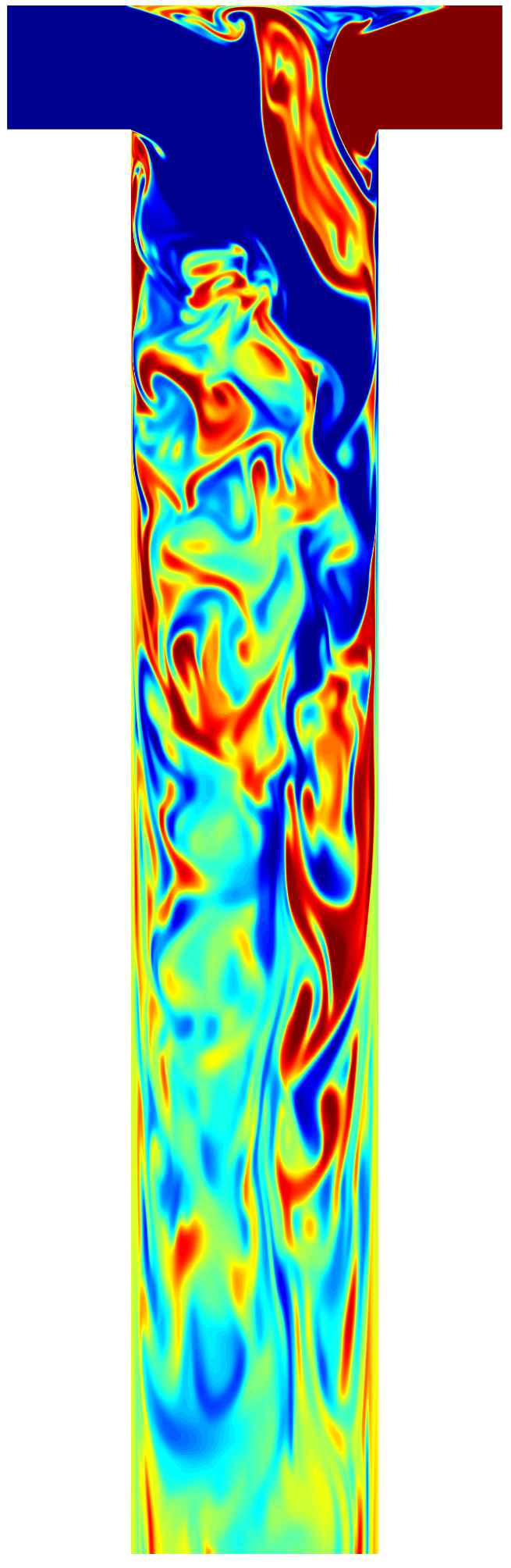} &
\includegraphics[width=0.12\textwidth]{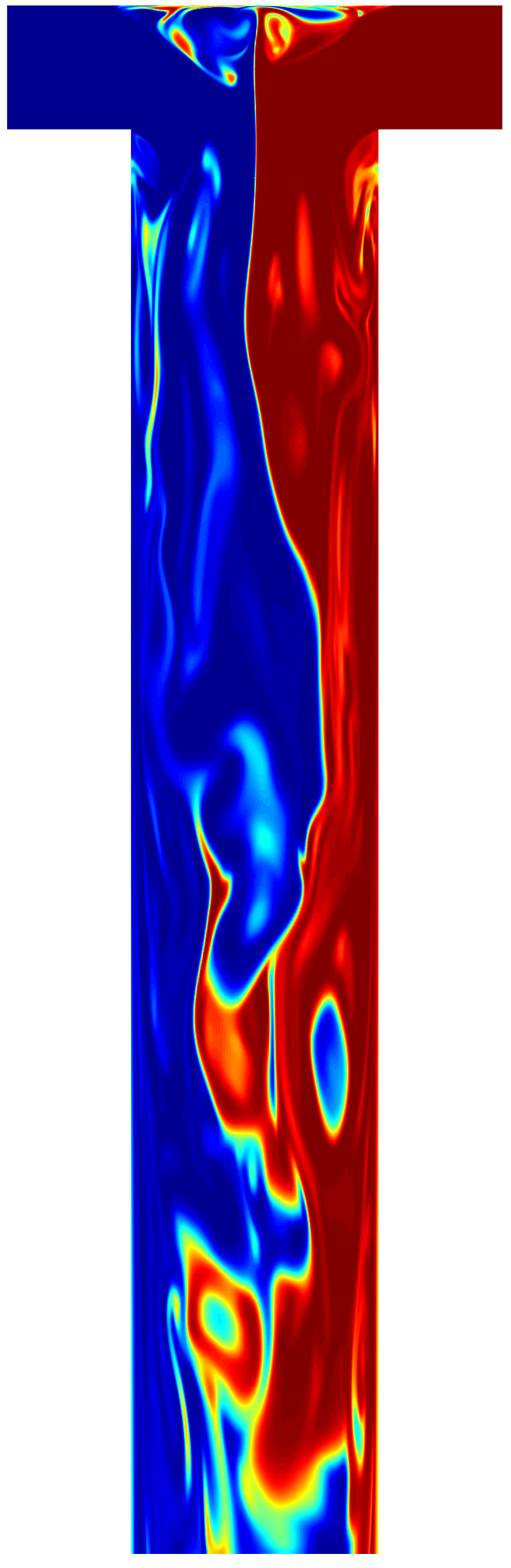} &
\includegraphics[width=0.12\textwidth]{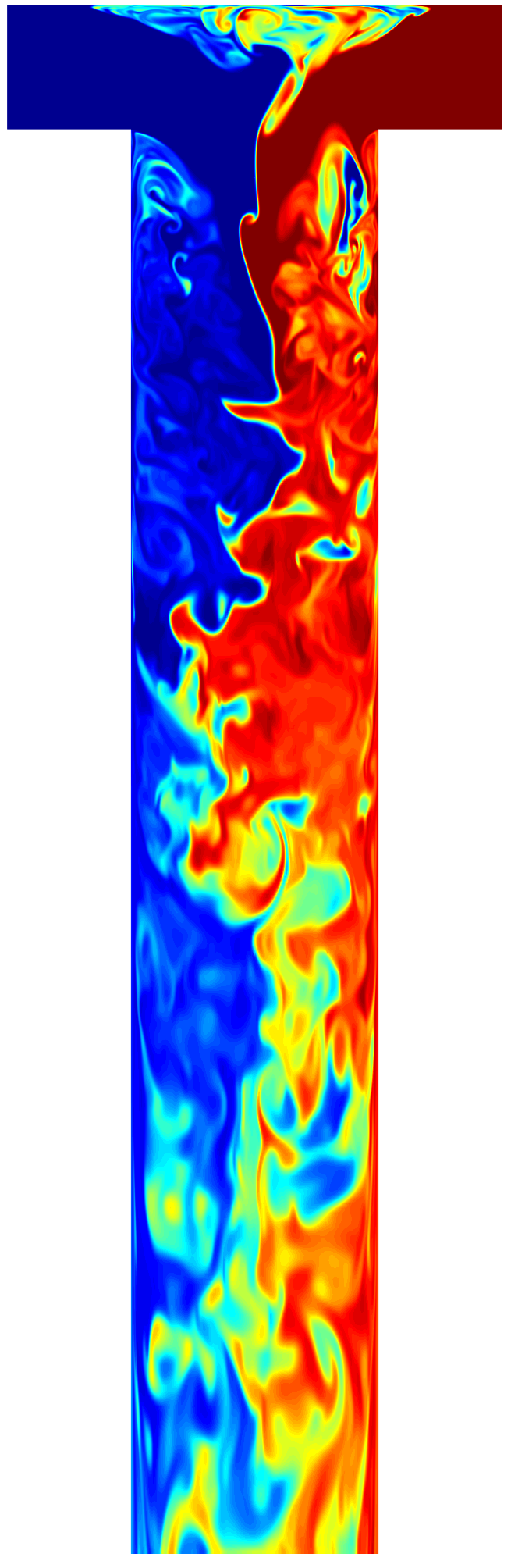} &
\includegraphics[width=0.12\textwidth]{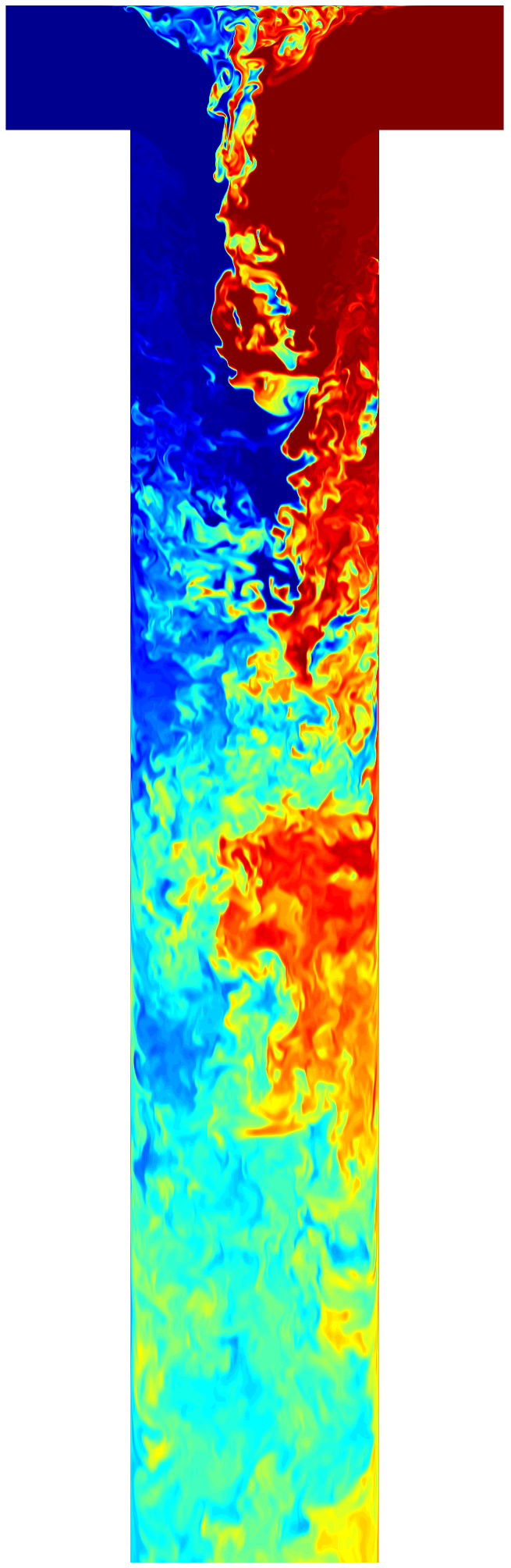}\\
\multicolumn{2}{c}{low Reynolds number regime}&\multicolumn{2}{c}{intermediate regime}&\multicolumn{2}{c}{turbulent regime}
\end{tabular}
\caption{(Color online) (a) T-mixer showing an instantaneous snapshot of passive scalar $\phi$ at $Re=200$, illustrated by four $(y,z)$-cross sections and one $(x,y)$-cross section which hereafter are used to visualize the flow and passive scalar behavior. (b) Experimentally measured mixing time $t_m$  as function of Reynolds number $Re$. Error bars indicate the standard deviation of $t_m$ which is extracted from a sample of at least three measurements. (c)--(h) Instantaneous snapshots of passive scalar $\phi$ at cross section $z=-0.25$ illustrating the mixing regimes considered within this work. }
\label{fig:micro_exp_tmix}
\end{center}
\end{figure*}

The quality of mixing determines the efficiency of many chemical reactions in fluids. Examples are chemical synthesis \cite{kim2016submillisecond,jensen2017flow,plutschack2017hitchhiker} and particle synthesis, such as fast precipitation of inorganic \cite{schwarzer2002experimental} and organic compounds \cite{johnson2003mechanism,karnik2008microfluidic}. At the interface of two mixing fluids, mass is exchanged between both fluids by diffusion, a process usually referred to as \emph{micromixing} \cite{baldyga1999turbulent,bockhorn2010micro}. The rate of mass transfer is quantified by the diffusion flux, which is the product of the area of the fluid interface and the diffusion coefficient $D$. As the diffusion coefficient is very small in liquid mixtures ($D\sim10^{-9}m^2/s$), the key in accelerating mixing is the enlargement of the fluid interface via vortices in turbulent flow, termed \emph{macromixing}. Hence sophisticated designs have been proposed to enhance turbulence and thereby optimize mixing \cite{nagasawa2005design}. However, the commonly accepted view is that the specific power input is the only effective parameter, i.e.\ improvements in mixing efficiency due to geometrical optimizations are associated with increased energy dissipation\cite{falk2010performance}. In this paper, we provide direct experimental and numerical evidence that manipulation of the inflow conditions can significantly improve the mixing performance without increasing the energy consumption.

We consider a water-water mixture in the T-mixer shown in Fig.~\ref{fig:micro_exp_tmix}(a) because of its simplicity, which allows a better understanding of the underlying physicochemical mechanisms. The main control parameter is the power input, which depends on the Reynolds number $Re=u_0 d/\nu$, where $d$ is the hydraulic diameter of the inlet, $u_0$ the mean inlet velocity and $\nu$ is the kinematic viscosity of the fluid. The Schmidt number $Sc=\nu/D$ sets the ratio of mass to momentum diffusion and is also an important parameter in mixing (in liquids, $Sc=\mathcal{O}(10^3)$). Steady and unsteady low $Re$ regimes have been intensively investigated in recent years by means of simulation and experiments with good agreement \cite{fani2013investigation,hoffmann2006experimental,bothe2006fluid,poole2013bifurcation,soleymani2008numerical,thomas2010experimental,fani2014unsteady,schikarski2017direct,mariotti2018steady}. Nevertheless, the turbulent mixing at $Re>2000$ has not been investigated so far, neither in experiments nor in simulations. This is surprising in view that in most applications the flow is well into the turbulent regime. 

In what follows, we present a detailed analysis of the mixing regimes encountered in the T-mixer as $Re$ increases up to  $4000$, see Fig.~\ref{fig:micro_exp_tmix}(b)--(h). We solve the Navier--Stokes equations using direct numerical simulation (DNS), whereby all scales of fluid motion are resolved in space and time \cite{moin1998direct}. This method stands out because it is free of empirical parameters, which in more commonly applied approaches such as RANS or plug-flow reactor models are needed to model turbulence . However, DNS remains mainly a research tool because of its high computing cost. For example, our DNS at $Re=4000$ required 1.4 million cores hours at the cluster MEGGIE of the Erlangen Regional Computing Center (RRZE) in order to obtain converged statistics.

In our experiments, we use the competitive-parallel Villermaux--Dushman reaction to determine the mixing time\cite{commenge2011villermaux}. By adapting the concentration of the chemicals and transforming the measured quantities to concentration-independent mixing times, we characterize mixing from the laminar to the fully turbulent regime $100\le Re\le4000$.

\section{Methods}
\label{sec:methods}

Our T-mixer consists of two inlets of square crosssection  with hydraulic diameter $d=1$\si{mm}, which discharge in a mixing channel of $1$mm height, $2$mm width and $11.5$mm length. This results in identical mean velocity ($u_0$) for the inlets and the mixing channel. In the experiment, the inlets are $18$mm long to allow the flow profiles to develop before entering the mixing channel. Note however, that the dimensionless length to achieve a fully developed laminar flow profile is $L/d=0.05 Re$. Hence, in our experiments fully developed laminar flow could only be achieved up to $Re=360$. By contrast, fully developed turbulent profiles require $L/d=4.4Re^{1/6}$ only\cite{beavers1970experiments,cengel2010fluid}, which was always satisfied in our experiments. In the numerical simulations, fully developed flows were imposed at the inlet boundary. This allowed using inlets of just  $3$mm in length, drastically reducing the computational effort.

\subsection{Numerical model}\label{sec:num}

We use Cartesian coordinates $x^*$, $y^*$, $z^*$ with the origin located in the center of the junction, $y^*$ defined along the opposing inlet channels and $x^*$ defined along the streamwise direction of the outlet channel. Lengths are made dimensionless by scaling with the hydraulic diameter, $x=x^*/d$,  $y=y^*/d$, $z=z^*/d$, whereas velocities are made dimensionless with $u_0$. The mixing process is governed by the dimensionless incompressible Navier--Stokes equations 
\begin{equation}
\frac{\partial \vec{u}}{\partial t} + \vec{u} \cdot \nabla\vec{u} = -\nabla p + \dfrac{1}{Re}\Delta \vec{u},\qquad \nabla \cdot \vec{u}= 0,
\label{eq:ns}
\end{equation}
coupled a the convection-diffusion equation for the concentration
\begin{equation}
\frac{\partial \phi}{\partial t} + \vec{u} \cdot \nabla\phi = \dfrac{1}{Re\,Sc}\Delta  \phi,
\label{eq:cd}
\end{equation}
where $p$ is the pressure, $\phi \in [0,1]$ is a passive scalar and $\vec{u}=(u,v,w)$ is the velocity field.

\subsubsection{Spatial and temporal resolution of mixing processes}

We discretized eq.~\eqref{eq:ns} with the finite-volume method in space and an explicit low storage Runge--Kutta scheme in time, both of 2$^{nd}$ order, by using the high-performance-code FASTEST3D \cite{schikarski2017direct}. Because our simulations solved for all fluid (eddying) motions in space and time, macromixing was simulated with very high fidelity. To ensure this, the criterion $h\lesssim 2\eta$ was applied, where $h=\text{max}(\Delta x, \Delta y,\Delta z)$ is the local grid spacing and $\eta$ is the smallest (Kolmogorov) flow scale. It is defined as $\eta = (\nu^3/\epsilon)^{1/4}$, where $\epsilon$ is the dissipation of the turbulent kinetic energy. This resulted in almost 200 million grid points for the highest Reynolds number considered ($Re=4000$). Furthermore, to enforce the stability of our explicit Runge-Kutta scheme, a constant time step was chosen such that the CFL number $c= |u/\Delta x| +  |v/\Delta y| + |w/\Delta z|<0.3$.

The smallest length scale of the scalar field is termed Batchelor scale ($\eta_b$) and is related to the Kolmogorov scale via the Schmidt number, $\eta_b = \eta/\sqrt{Sc}$. In the range of scales between the Kolmogorov and the Batchelor scales, micromixing takes place. Here the velocity field features no fluctuations, whereas the scalar field fluctuates. Although micromixing has been sometimes argued to be the rate-limiting step in liquid mixing\cite{fournier1996newex}, it cannot be resolved in the simulations because of $Sc=\mathcal{O}(10^3)$. In practice, spatial discretization schemes and subgrid modeling introduce a further diffusion term in Eq.~\ref{eq:cd}, with a corresponding artificial diffusion coefficient. This artificial diffusion coefficient depends on the local grid size, numerical scheme and subgrid model, and is in general larger than the natural diffusion coefficient. 

In our simulations, the Schmidt number is $Sc=600$, so resolving the Batchelor scale would require over $6\times10^{12}$ grid points at $Re=4000$. As this is computationally infeasible, we treat the convective term in Eq.~\ref{eq:cd} to ensure a bounded value of $\phi$ and a nearly 2$^\text{nd}$ order approximation despite the high Peclet numbers $Pe=Re\,Sc$. This is accomplished by using linear schemes with flux limiters fulfilling the total variation diminishing criteria \cite{waterson2007design}. In a  nutshell, our scheme essentially adds diffusion (locally in space and time), when large gradients of $\phi$ appear. A complete description of the numerical technique used herein and a thorough validation is given in \citet{schikarski2017direct}. 

\subsubsection{Boundary conditions: flow regimes in a square duct}

\begin{figure}[h]
	\begin{center}
        \includegraphics[width=0.48\textwidth]{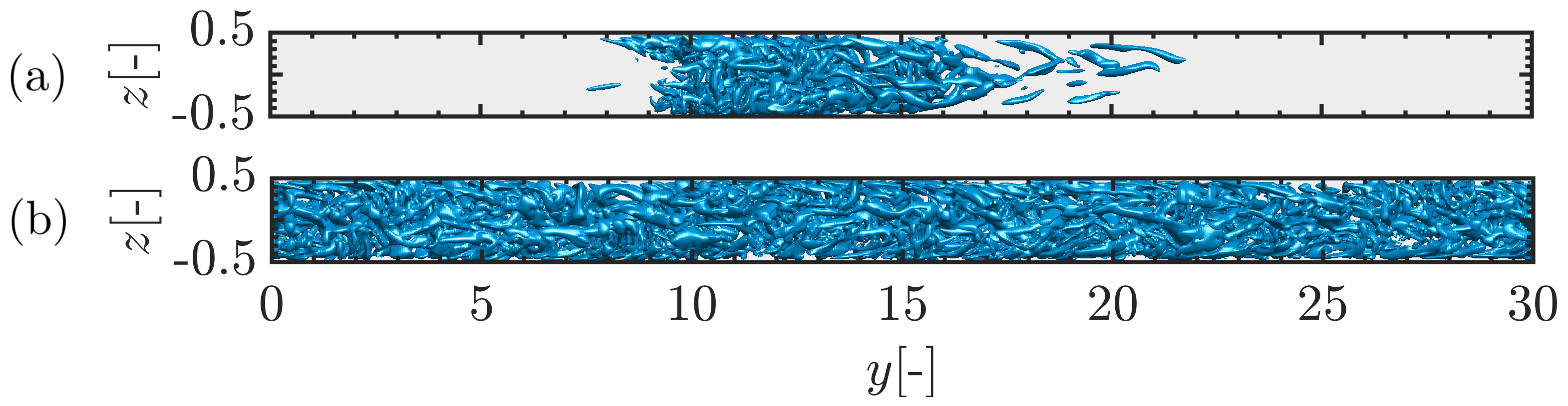}
		\caption{Turbulent flow in a square duct driven by a constant volume flux. The boundary conditions are periodic in the streamwise direction in a domain of length 30$d$. (a) Localized turbulent puff at $Re=1700$. (b) Fully turbulent flow at $Re=2500$ . Instantaneous isosurfaces of normalized $Q$-criteria with threshold $Q=1$  \cite{jeong1995identification}.} 
		\label{fig:puff}
	\end{center}
\end{figure}

Because of the wide range of $90\le Re\le 4000$ covered in our simulations, special care must be taken with the inlet boundary conditions.  At low $Re$, the flow in a square duct is always laminar. In fact, laminar square-duct flow is linearly stable at all $Re$ and in DNS it is necessary to disturb the flow in order to generate turbulence via nonlinearities. In the presence of strong disturbances, localized turbulent puffs were observed in experiments \cite{delozar2010universality} and numerical simulations\cite{takeishi2015localized} for $Re \gtrsim 1430$, see Fig.~\ref{fig:puff}(a). In fact, Osborne Reynolds had  already observed such puffs in his classical experiments of cylindrical pipe flow\cite{Reynolds83}. As $Re$ increases, puffs occur with increased probability \cite{wygnanski1973,avila2013}, until the puff regime is replaced by fully turbulent flow at $Re\gtrsim 2030$ \cite{barkley2015}. A snapshot of fully turbulent flow at $Re=2500$ is shown in Fig.~\ref{fig:puff}(b).

A full understanding of the dynamics of transition in pipes and ducts has just emerged in the last decade. The interested reader is referred to \citet{Barkley16} for a recent comprehensive review. Here, we briefly summarize the main point for engineering purposes. For $Re \gtrsim 1430$ the flow in a square duct can either be laminar or turbulent depending on the level of disturbance. For example, Osborne Reynolds could keep the flow laminar up to  $Re\approx 12,000$ with a purposely designed entrance and with the fluid settled in the supply tank \cite{Reynolds83}.  However, in conventional setups used in chemical engineering like ours, the occurrence of turbulent puffs, separated by segments of laminar flow, is a random process controlled by imperfections and natural disturbances. This \emph{natural} onset of turbulence occurred in our experiments at $Re\approx 1400$, as shown in Fig.~S1 of the ESI\dag. Finally, it is worth noting that at a given Reynolds number, the flow in the T-mixer is much more turbulent than in a square-duct. For example, the level of turbulence reached near the junction at $Re=4000$ is similar to that of cylindrical pipe flow at $Re\approx 6\times10^4$ (see S2 in ESI\dag).

\begin{figure}
\begin{center}
\includegraphics[width=0.44\textwidth]{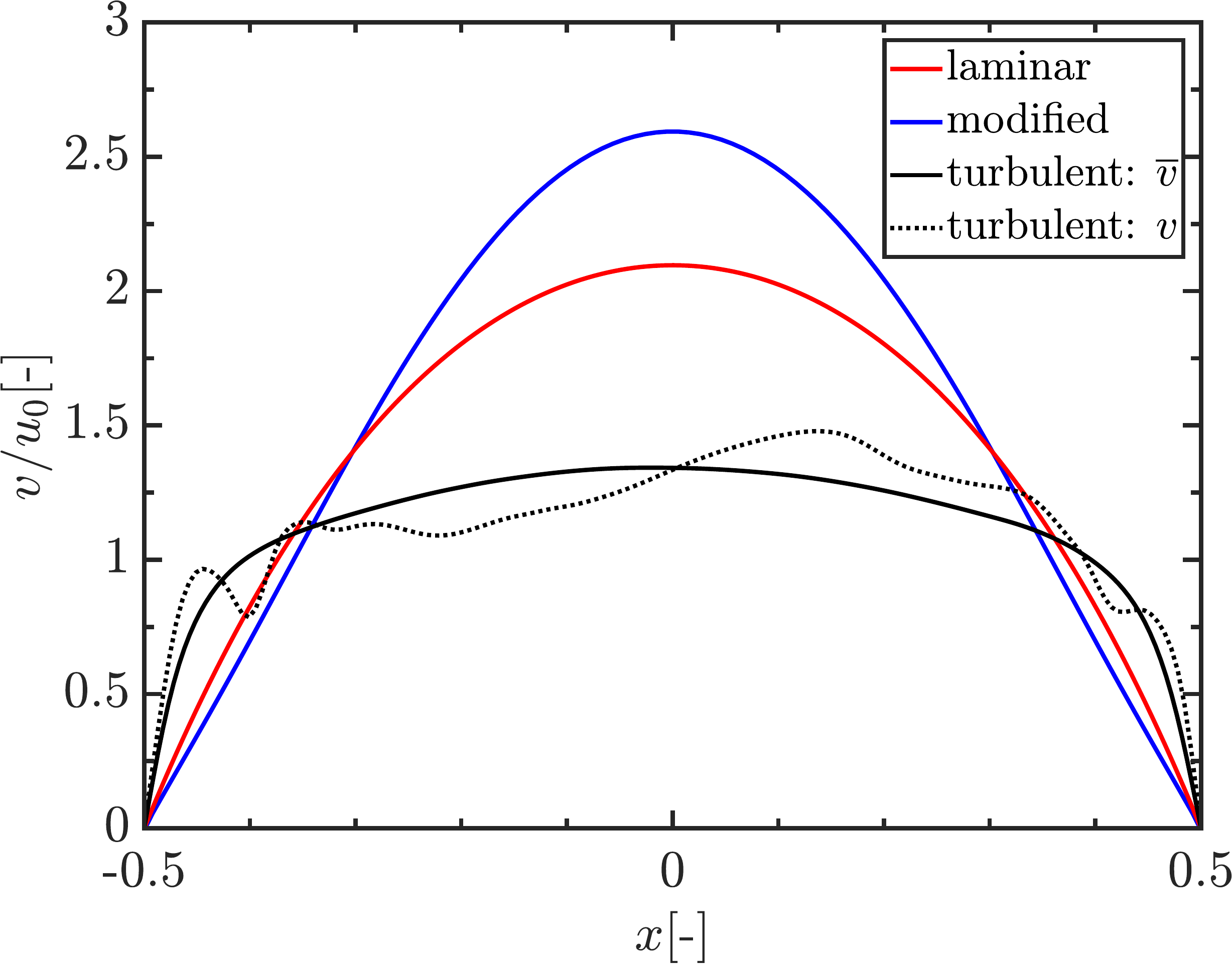}
\caption{Inlet velocity profiles $v$ at mid-height ($z=0$) used as boundary condition for the left inflow channel ($y<0$). The laminar and modified laminar inflow velocities are steady, whereas for turbulent inflow the boundary conditions are time-dependent. An instantaneous profile $v$ and the time-averaged profile $\overline{v}$ at $Re=4000$ are shown here for clarity.}
\label{fig:inflow_cross}
\end{center}
\end{figure}

In our numerical simulations, we tested the influence of the inflow conditions on the mixing process by imposing either fully developed turbulent or laminar flow at the inlets. Turbulent boundary conditions were generated by simulating the flow separately in a square duct with periodic boundary conditions in the streamwise direction. As shown in S3 of ESI\dag, our first and second order statistics of the square duct flow agree very well with references values from literature \cite{pinelli2010reynolds, gavrilakis1992numerical}. The simulated data were stored over time at a given cross-section and then prescribed at the inlet during the simulation of the T-mixer. Fig.~\ref{fig:inflow_cross} shows the inlet streamwise velocity $v$ over $x$ at mid height for the laminar velocity profile (red, $Re$-independent in dimensionless form), the time-averaged 
and an instantaneous turbulent velocity profile (black) at $Re=4000$. In addition, a modified laminar flow (blue) preserving the cross-sectional volume flux is shown. This was used to test the sensitivity of the dynamics observed in the intermediate regime to inflow boundary conditions (see \S\ref{sec:intermediate}).

\subsubsection{Intensity of segregation}

All the statistical data presented in the paper were obtained over a time interval of $200d/u_0$ following a transient of $40d/u_0$, over which data is not collected. We quantified the mixing efficiency with the intensity of segregation\cite{danckwerts1952definition}
\begin{equation}
I_s=\frac{\sigma_b}{\sigma_\text{max}},
\label{eqn:dom}
\end{equation}
which is $1$ ($0$) for completely segregated (perfectly mixed) streams. Here $\sigma_\text{max}$ is the maximum variance (determined by completely segregated streams),
 \begin{equation}
\sigma^2_\text{max} = \langle \phi\rangle(1-\langle \phi\rangle) 
\end{equation}
and $\sigma_{b}$ is the mean square deviation of the concentration profile 
\begin{equation}
 \sigma^2_{b} = \frac{1}{S_\text{tot}} \int_S (\phi-\langle \phi\rangle)^2 \, \mathrm{d}y\, \mathrm{d}z 
=  \frac{1}{S_\text{tot}} \sum_{i=1}^N S_i (\phi_i-\langle \phi\rangle)^2.
\end{equation}
The mean concentration is 
\[
\langle \phi\rangle = \dfrac{1}{S_\text{tot}}\sum_{i=1}^N S_i \,\phi_i,
\]
where $S_\text{tot}$ ($S_{i}$) is the area of the cross section (control volume $i$) and $N$ the number of control volumes in the cross-section. 

\subsection{Experimental methods}
\label{sec:experimental_methodology}

The main channel of the T-mixer and the inlets were milled into a piece of stainless steel and covered by a plate of the same material that was cold-welded on top. Both inlets were connected to in-house build piston-pumps with a volume of 50 ml each. Both pistons were moved simultaneously in their glass-barrels by a stepping motor with gear reduction enabling a pulsation-free flow. The pressure at each inlet was monitored with a sampling rate of 1 kHz using pressure transmitters (model A09 from Sensor-Technik Wiedemann GmbH, Germany). A LabView2009 program was developed to control the pistons and to monitor the pressure. In order to be able to scan different Reynolds numbers within one experimental run, a program was developed to realize several consecutive flow rates without interruption. The temperature of the barrels was kept at $20\pm1^\circ$C using a LAUDA Alpha R8 cooling thermostat.

Pressure loss measurements are the first essential step in a comparative study between simulation and experiment. In our numerical simulations, pressure losses were measured from the inlets to the outlet. In the experiments, pressure probes were installed before the connectors to the inlets of the T-mixer to avoid disturbances to the fluid flow. As a result, the experimental measurements included also the losses due the connector and the additional $15$mm of inlet. Their contribution was measured independently in a separate set of experiments, in which the T-mixer was replaced by a square duct of $15$mm in length. This was subtracted from the total loss measured in the mixing experiments, thereby allowing a direct comparison to the simulations (see Fig.~S4 in ESI\dag\ for a sketch). Finally, we note that at low Reynolds numbers ($Re\lesssim400$) the pressure loss was of the order of the measurement uncertainty, hence large errors were expected in this regime.

Chemical reactions, used as molecular probes for mixing, are powerful tools to gain insights into the mixing process. We here used the common Villermaux--Dushman reaction\cite{commenge2011villermaux,falk2010performance} to quantify the mixing efficiency. Details of the employed chemical components, experimental protocols and data analysis are in S5 of ESI\dag.

\section{Mixing regimes}
 \begin{figure}[h]
\begin{center}	
\includegraphics[width=0.45\textwidth]{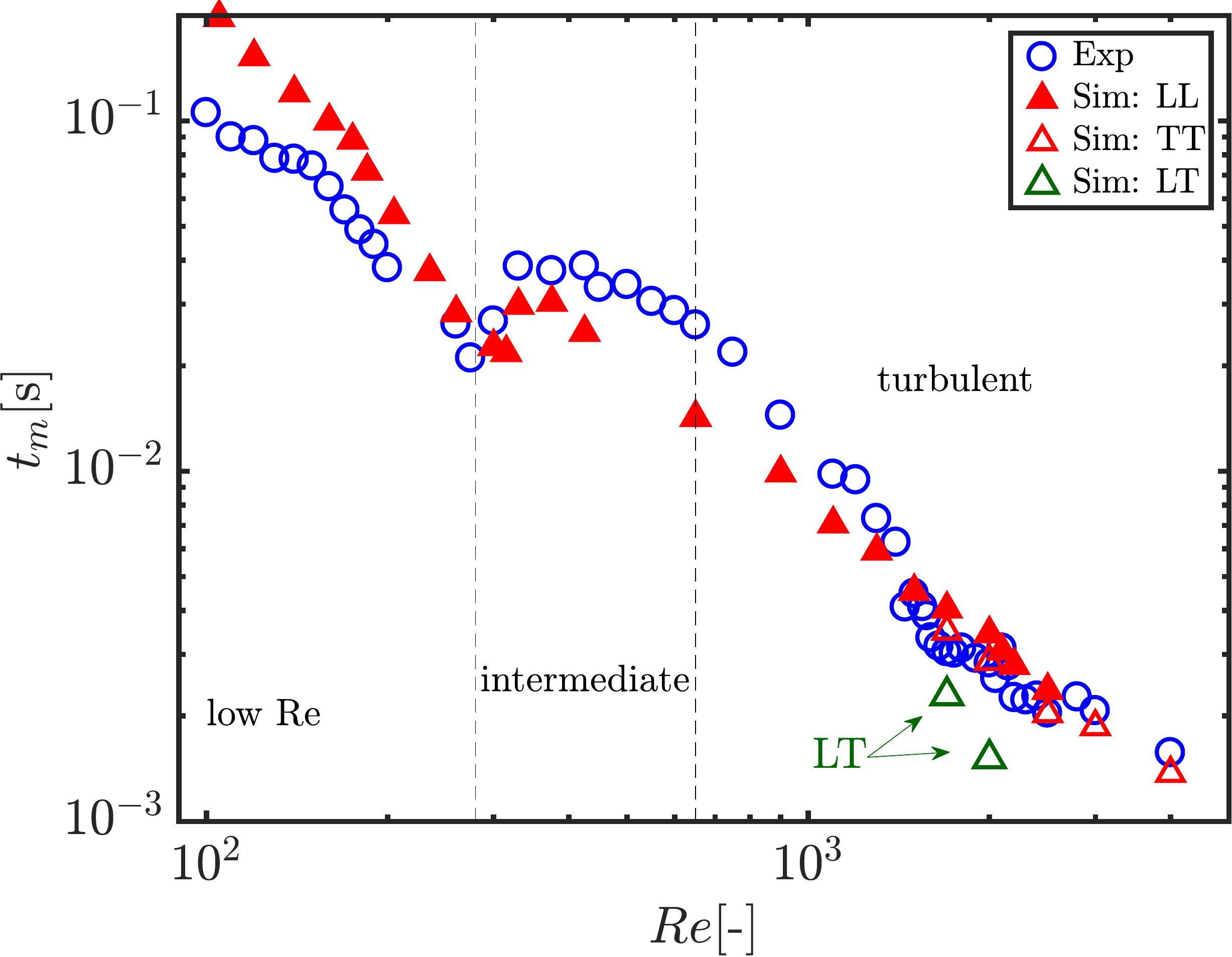}
\caption{Experimentally measured mixing time $t_m$ (circles) and computationally estimated mixing time using Eq.~\ref{eq:equality} (triangles) as a function of $Re$. Three mixing regimes are highlighted: \textit{low $Re$, intermediate and turbulent}. These are separately discussed in \S\ref{sec:lowRe}, \S\ref{sec:intermediate} and \S\ref{sec:turbulent}, respectively. $ \blacktriangle$ and $\triangle$ triangles represent laminar-laminar (LL) and turbulent-turbulent (TT) inflow conditions, respectively. Data with laminar-turbulent (LT) inflow conditions are indicated with an arrow. The influence of the different inflow conditions is discussed in \S\ref{sec:turbulent}.}
\label{fig:mixing_regime}
\end{center}
\end{figure}

The mixing efficiency is quantified in the Villermaux--Dushman characterization\cite{commenge2011villermaux,falk2010performance} with the mixing time $t_m$ (or normalized segregation index $X_s$), which is the time needed to achieve the final segregation between two competitive reaction products in relation to the initial feed concentration. By contrast, the mixing efficiency is quantified in the simulations with the dimensionless intensity of segregation $I_s$ (eq.~\ref{eqn:dom}), which measures the \emph{local} segregation at a particular point in the mixing device (here the outlet). In order to allow a comparison between simulation and experiment, the intensity of segregation is multiplied here by the mean residence time from the junction to the outlet $t_r=12d/u_0$. We find that the simple empirical relationship
\begin{equation}
t_m = 1.8\,I_s\cdot t_r,
\label{eq:equality}
\end{equation}
renders excellent agreement between the numerically and experimentally determined mixing time, especially in the high $Re$-regime ($Re \gtrsim1500$). The empirical factor $1.8$ was chosen here such that the experimental $t_m$ and the numerical $I_s\cdot t_r$ match exactly at $Re=2500$. This particular Reynolds number was selected to avoid the hydrodynamic complexities of the transitional regime ($Re<2000$) and to ensure that fully developed flow was present in the inlet channels of the T-mixer. We note that in the Villermaux--Dushman characterization the conversion from the segregation index $X_s$ to mixing time is accomplished through an empirical constant as well (see Fig.~S5 in ESI\dag). The qualitative changes observed in the mixing efficiency as $Re$ increases are elucidated in what follows.

\subsection{Low Reynolds number regime}\label{sec:lowRe}

 \begin{figure}
\begin{center}	
\includegraphics[width=0.4\textwidth]{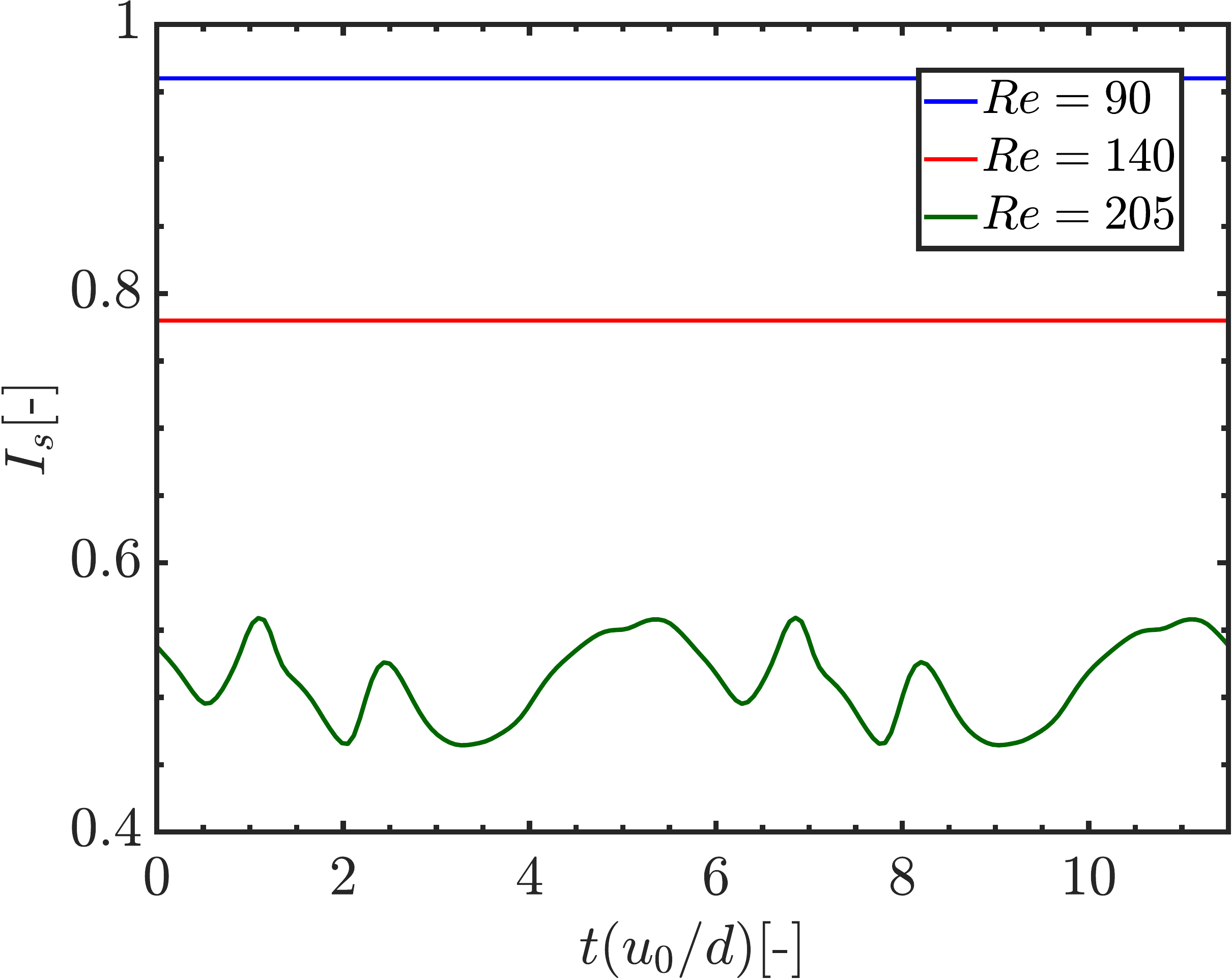}
\caption{Temporal evolution of the segregation index $I_s$ at $x = 11.5$ for various Reynolds numbers.}
\label{fig:lowRe_regime}
\end{center}
\end{figure}

When the Reynolds number is arbitrarily small (Stokes flow), the flow has no vortices (\emph{stratified} regime), and the inlet streams collide and flow side by side along the mixing channel. As $Re$ increases, four symmetric vortices gradually arise at the junction, but the planar interface persists, as shown generically by the colormap of $\phi$ in Fig.~\ref{fig:micro_exp_tmix}(c). Because the flow is left-right symmetric in both the stratified and the vortex regimes, mixing occurs only via molecular diffusion at the planar interface, whose width grows approximately as $\sqrt{2Dx/u_0}$ \cite{tabeling2005introduction}. In our simulations, this vortex regime was found stable up to $Re_c\approx 105$, at which point it was superseded by two counter-rotating vortices which break the reflection symmetry. They engulf and rotate the inlet streams along the main channel, resulting in an increase of mixing efficiency due to the enlarged interface area, see Fig.~\ref{fig:micro_exp_tmix}(d).  The \emph{engulfment} regime is a robust feature of T-mixers, but its onset depends slightly on the he height-to-width aspect ratio of both inlet and outlet channels \cite{poole2013bifurcation}. As $Re$ was further increased, the flow began to pulsate periodically in time as shown in Fig.~\ref{fig:lowRe_regime}. This marked the transition from the steady to the time-periodic engulfment regime, which was observed at $Re\approx175$ in the simulations and at $Re\approx160$ in the experiments. The two engulfing vortices remain, but the locations of their cores pulsate periodically with time. This unsteadiness enhances mixing, as seen in the change of slope of $t_m$ in Fig.~\ref{fig:mixing_regime}.

The experimentally measured mixing time $t_m$ is lower than that estimated numerically. This can be explained as follows. At low $Re$, the outlet channel was too short for the chemical reaction to complete, see \eg  the segregated streams leaving the outlet channel in Fig.~\ref{fig:micro_exp_tmix}(d). We thus suspect that here mixing was enhanced as the fluids steadily dropped in the beaker, and that the fluids continued to mix in the beaker, before being measured off-line in a UV spectrometer (see S5 of ESI\dag). For example, for $Re<80$ no experimental measurements were conducted because in the steady vortex regime the mixing time was always determined by the ongoing mixing in the beaker. As $Re$ was increased, the fluids mixed more intensively in the main channel and the ongoing mixing in the beaker became (relatively) less important. Hence, progressively better agreement between simulation and experiment was achieved by increasing $Re$. 

\subsection{Intermediate regime}\label{sec:intermediate}
\begin{figure}[h]
\centering
(a)\\\includegraphics[width=0.44\textwidth]{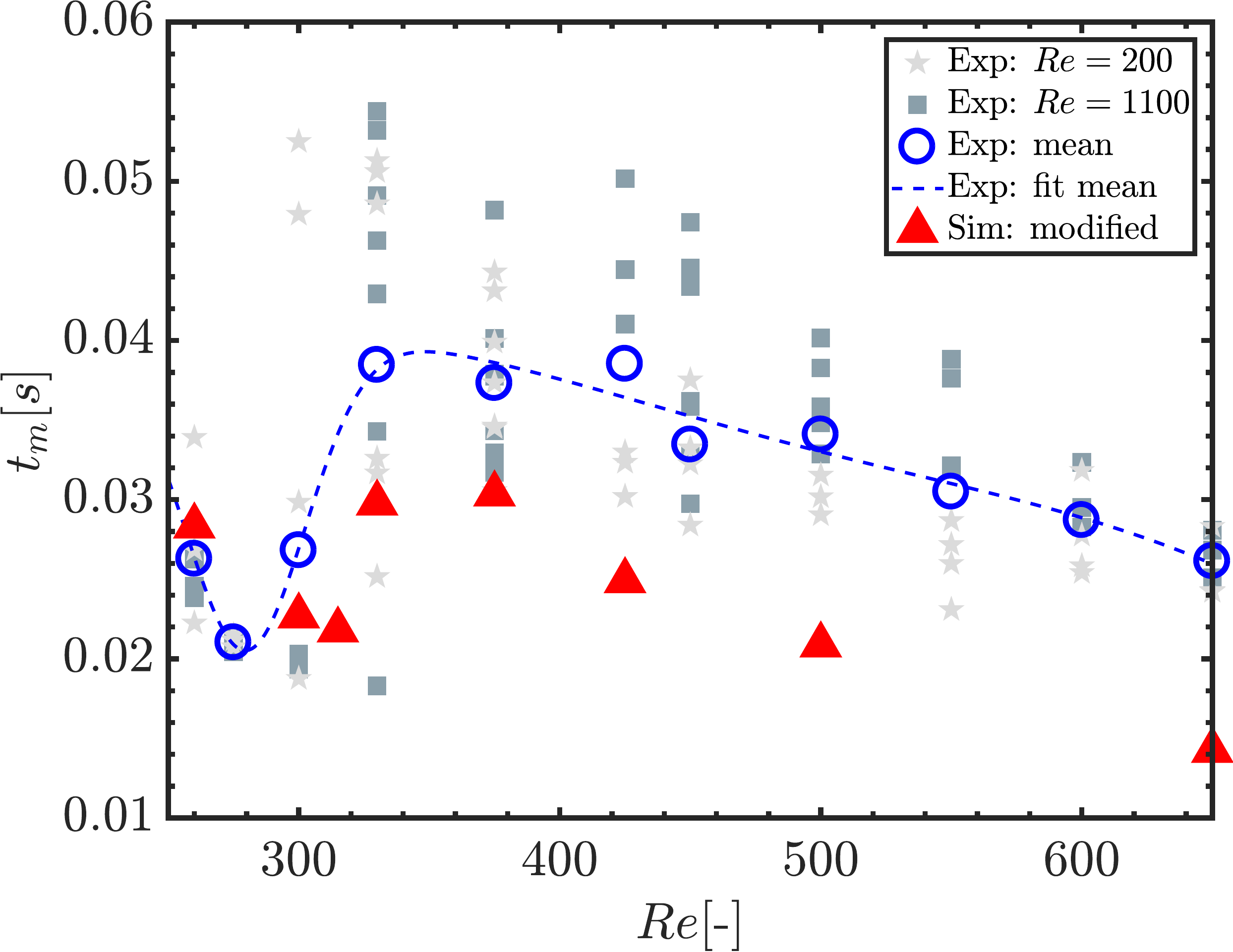}
\\(b)\\\includegraphics[width=0.40\textwidth]{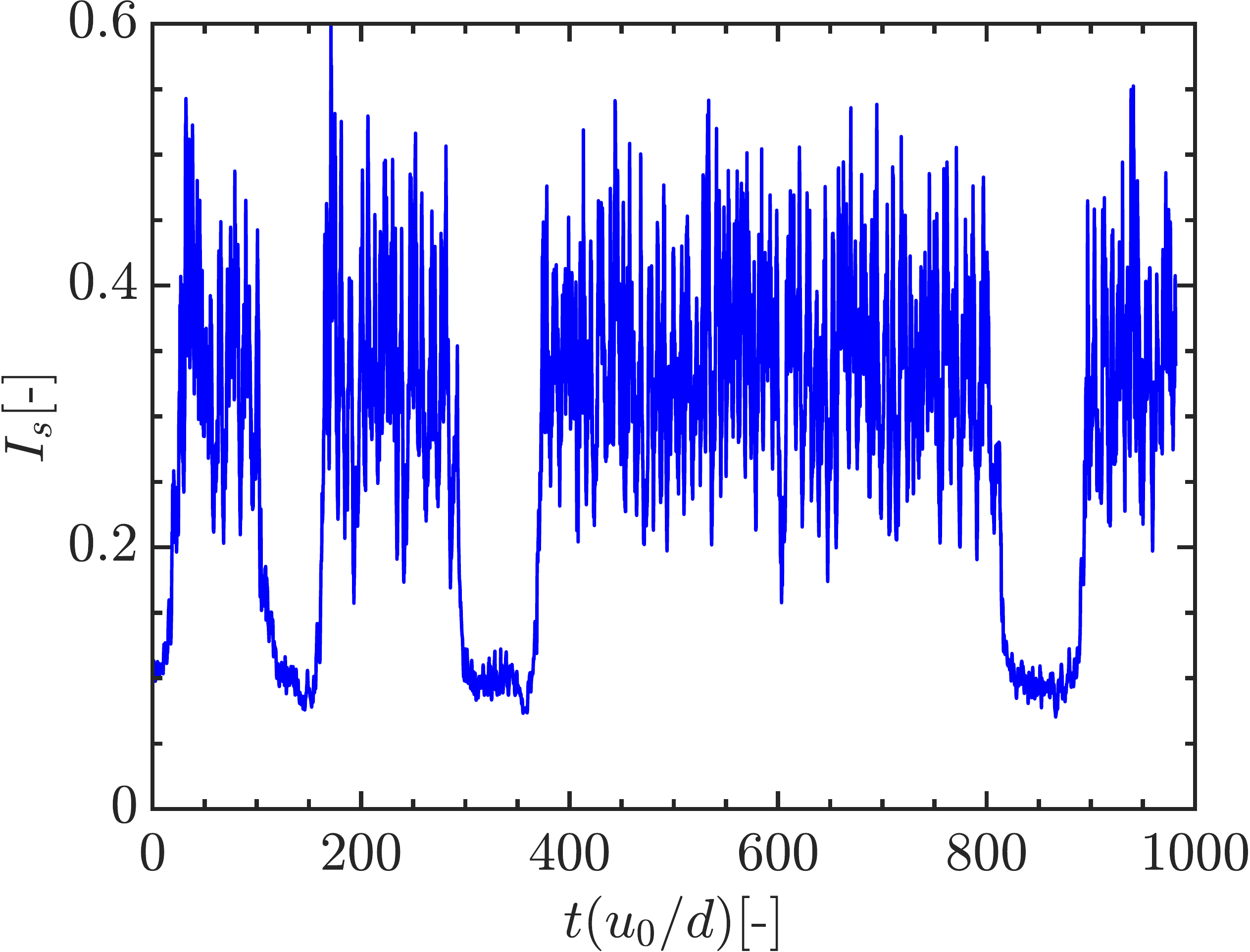}
\caption{(a) Experimentally measured mixing time $t_m$ (blank circles) and computationally estimated mixing time using Eq.~\ref{eq:equality} (triangles) as function of Reynolds number $Re$. Square and diamond symbols represent $t_m$ of experimental run using initial conditions at $Re=200$ and $Re=1100$, respectively. (b) Time history of  $I_s$ evaluated shortly after the junction at $x=2.5$ for $Re=330$. In the DNS, the modified laminar inflow condition of Fig.~\ref{fig:inflow_cross} was used.}
\label{fig:micro_exp_tmix_hysteresis}
\end{figure}
As the Reynolds number is further increased, the engulfment regime is replaced by a four vortex state in which the time-averaged flow field is reflection symmetric. This transition exhibits hysteresis in the simulations and for example at $Re=330$ the two states (engulfment and symmetric)  can be obtained depending on how the system is initialized \cite{schikarski2017direct}. In the hysteresis region, we attempted to obtain both states in the experiments by using the following procedure. First, the system was run at $Re=200$ or $Re=1100$, where only the engulfment or the symmetric are stable, respectively. Then the flow rate was rapidly adjusted, without interruption of the fluid flow, to a target Reynolds number (\eg $Re=330$), where the two states should be stable according to the simulations. For each Reynolds number investigated in this intermediate regime ($260\lesssim Re \lesssim 600$), this procedure was repeated several times, but no hysteresis could be detected. Instead, a large variability in the measured mixing time $t_m$ was observed depending on the realization. Fig.~\ref{fig:micro_exp_tmix_hysteresis}(a) shows that when averaging over all realizations, a continuous transition from the engulfment to the symmetric regime is recovered. This suggests that either the slightly different inlet profile (which was not fully developed in the experiment for $Re>360$) or other experimental imperfections triggered dynamic transitions between the two states, resulting in very different $t_m$ depending on how much time the flow exhibited a symmetric or engulfment pattern in each experimental run. 
 
This hypothesis was verified in the numerical simulations by replacing the laminar inlet boundary condition with a modified velocity profile (see the blue line in Fig.~\ref{fig:inflow_cross}). With this modified inlet profile, the flow was observed to jump randomly in time between the two states, as shown in the time series of Fig.~\ref{fig:micro_exp_tmix_hysteresis}(b) and the flow snapshots of Fig~ \ref{fig:micro_exp_tmix}(e)--(f). Hence, modifying the inlet boundary conditions destroyed the hysteresis regime and the same continuous transition as in the experiments was observed in the simulations. Such phenomenology is well known in fluid dynamics \cite{zimmerman2011bi} and will be disseminated separately in detail.

\subsection{Transition to turbulence}\label{sec:turbulent}

\begin{figure}[h]
	\begin{center}
		\includegraphics[width=0.44\textwidth]{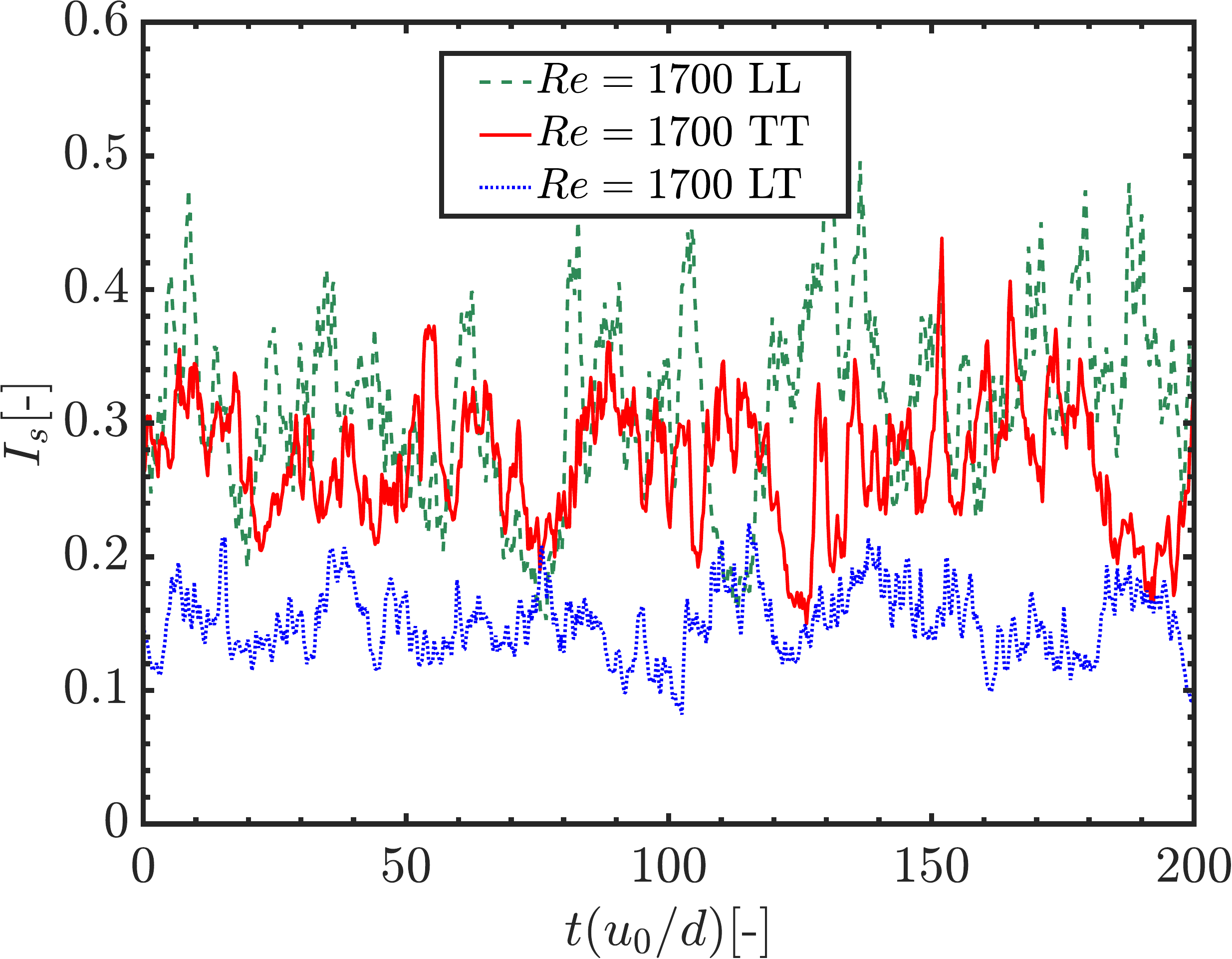}
				\caption{\label{fig:dom1700}Temporal evolution of intensity of segregation $I_s$ at $Re=1700$. Boundary conditions with  laminar-laminar (LL), turbulent-turbulent (TT) and laminar-turbulent (LT) inlets are compared. The temporal average of the time series  is $\overline{I_s}=0.32$, $0.27$ and $0.18$ for LL, TT and LT, respectively.}	
	\end{center}
\end{figure}

\begin{figure}[h]
	\begin{center}
		\begin{tabular}{cc}
			(a) & (b) \\
			\includegraphics[width=0.24\textwidth]{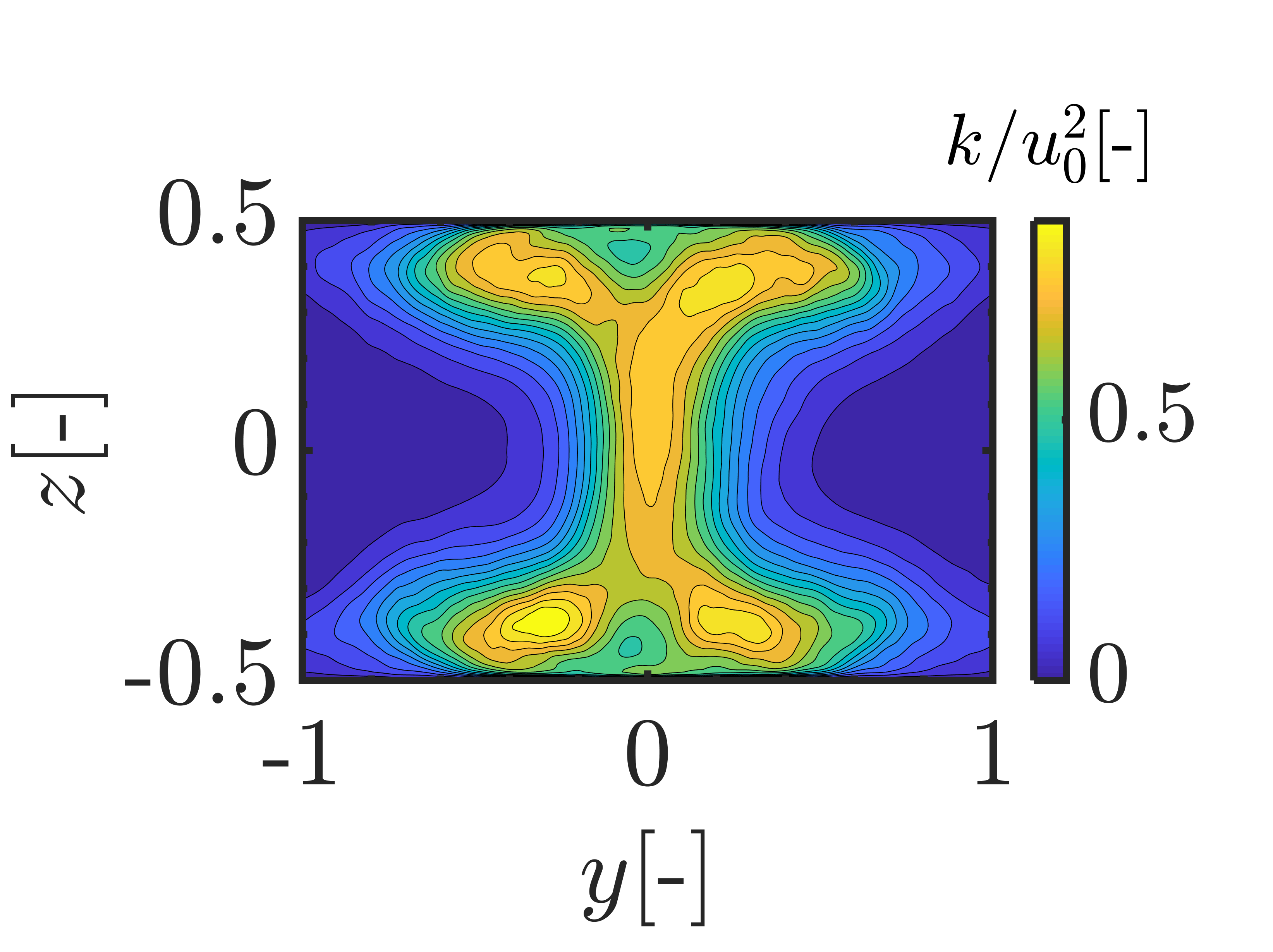} & \includegraphics[width=0.24\textwidth]{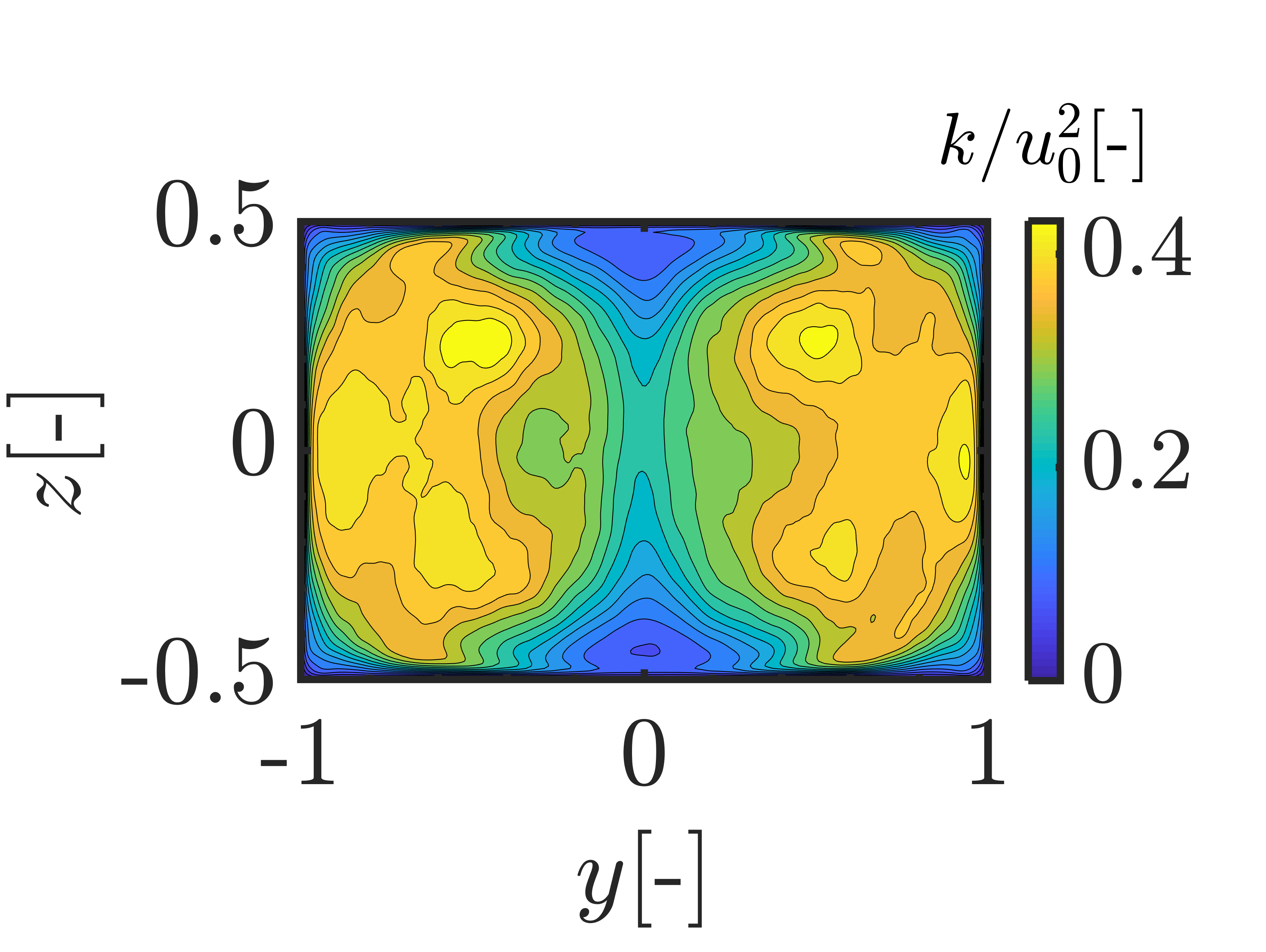} \\
			\includegraphics[width=0.24\textwidth]{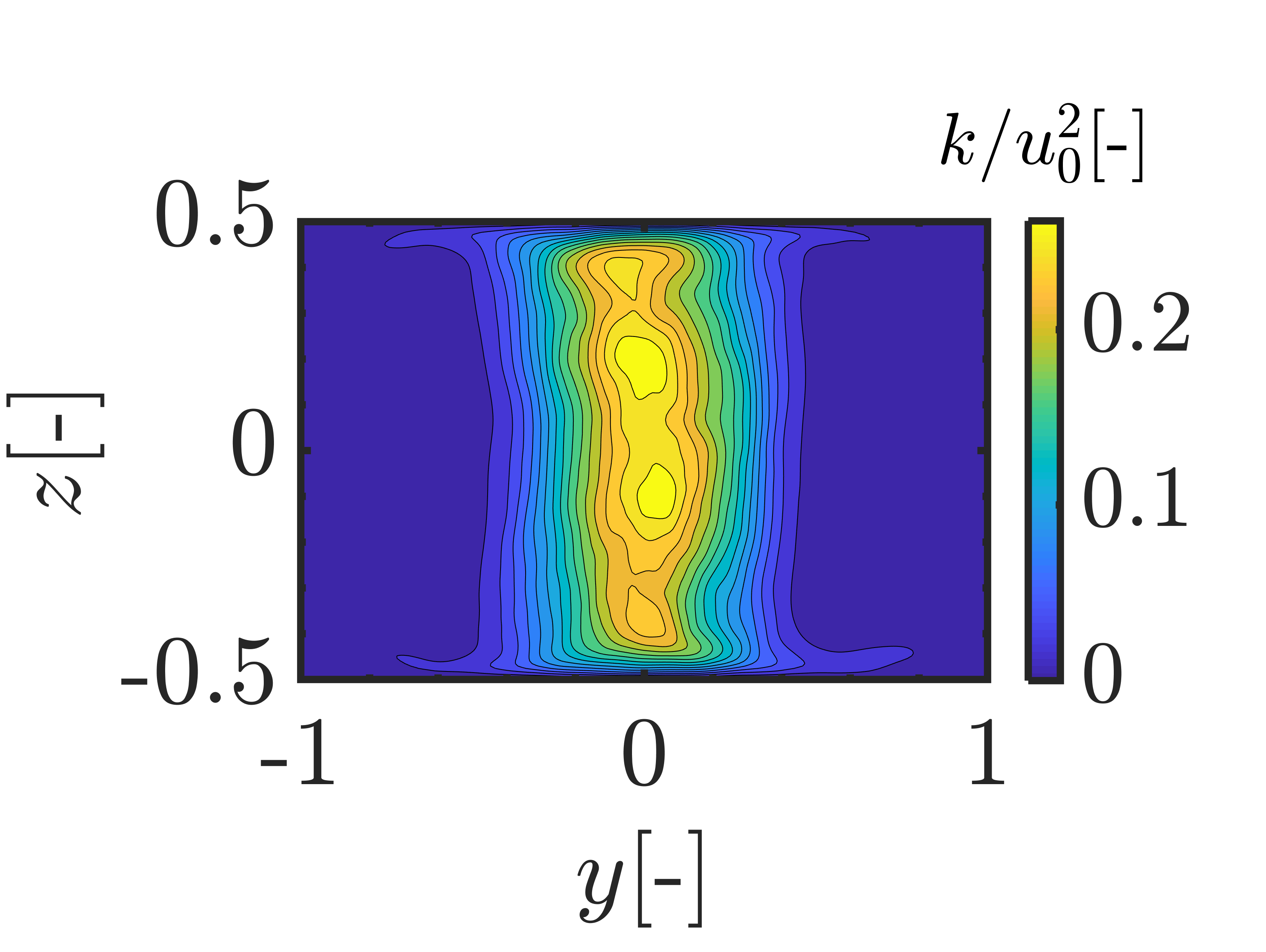} &\includegraphics[width=0.24\textwidth]{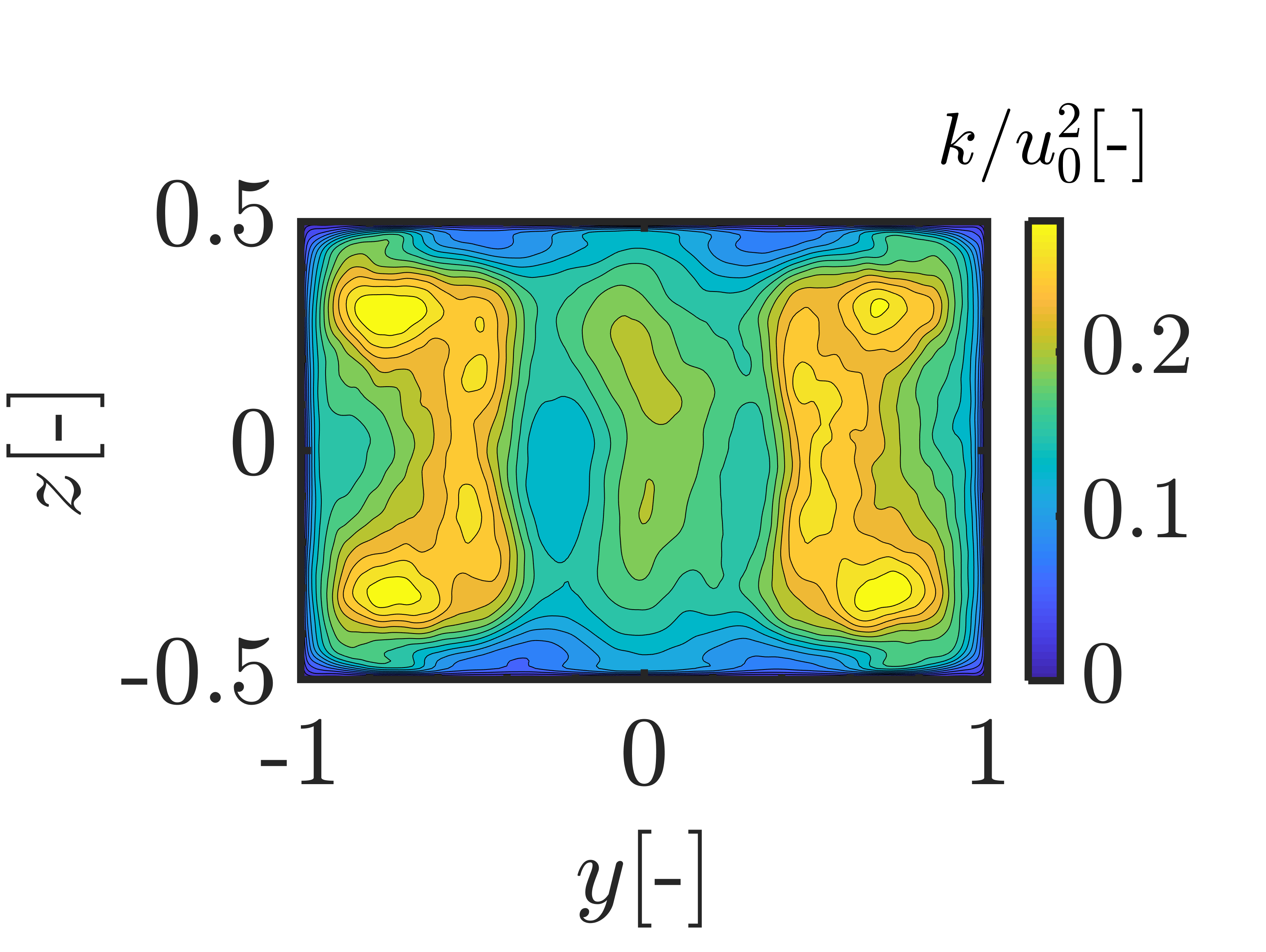}\\
			\includegraphics[width=0.24\textwidth]{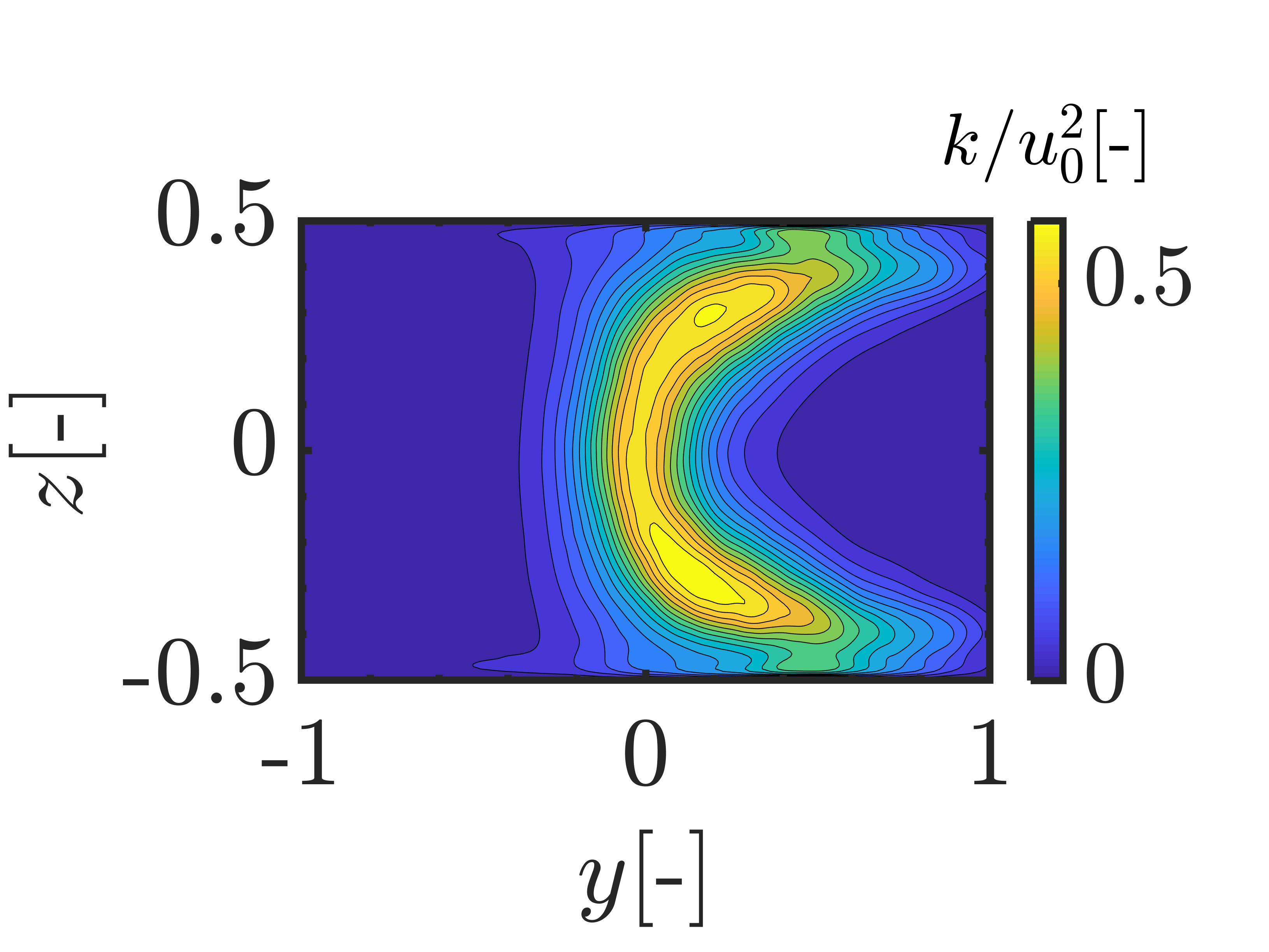} &
			\includegraphics[width=0.24\textwidth]{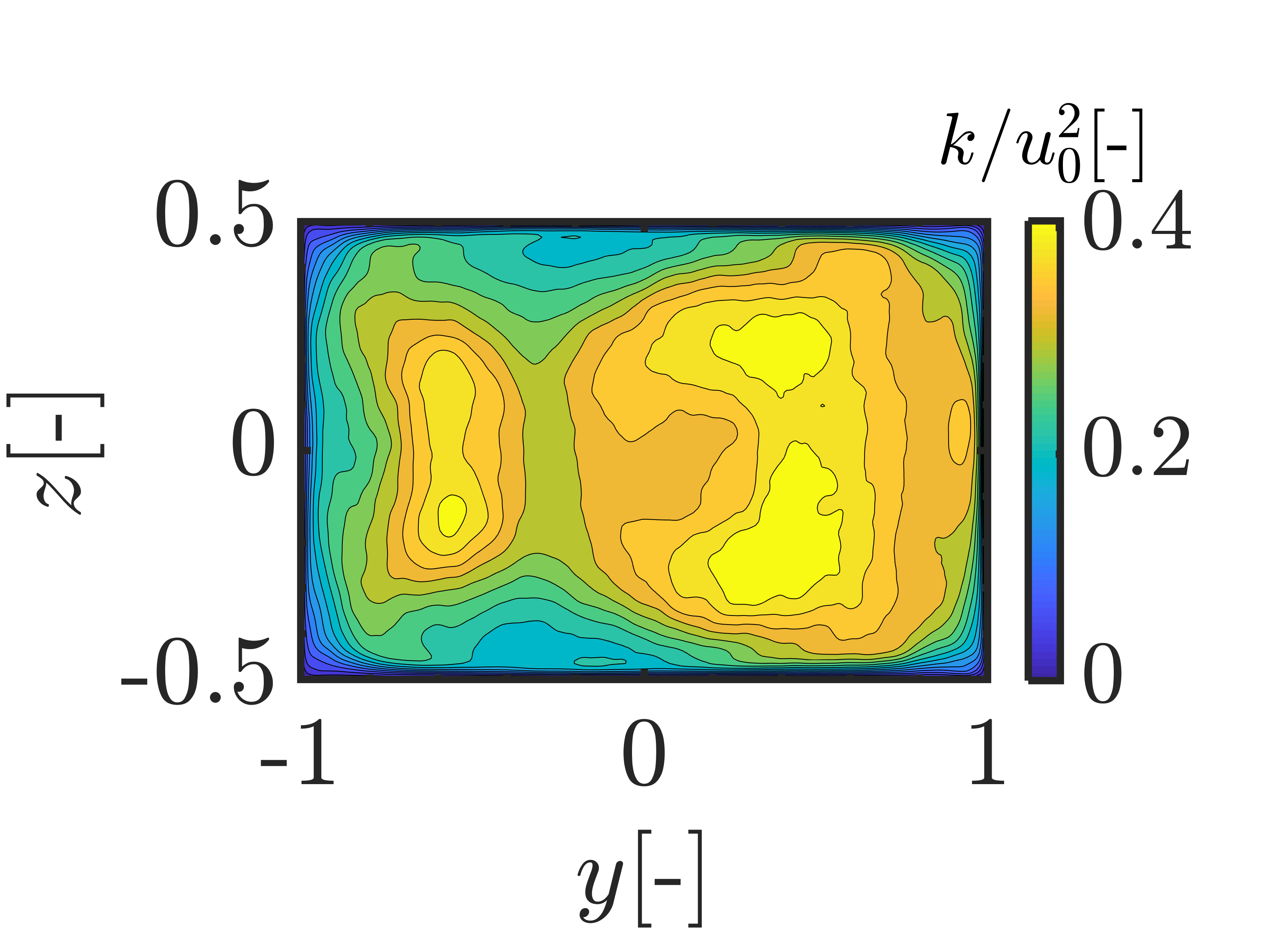}\\
		\end{tabular}
		\caption{Colormaps of normalized turbulent kinetic energy $k/u_0^2$ at two cross-sections, $x=0.25$ (a) and $x=1.5$ (b), along the mixing channel at $Re=1700$. From top to bottom laminar inflows (LL case),  turbulent inflows and  mixed inflows, consisting of turbulent and laminar inflow from the left side ($y<0$) and the right side ($y>0$), respectively, are shown.}
		\label{fig:tke_cross}
	\end{center}
\end{figure}

\begin{figure}[h]
	\begin{center}
\includegraphics[width=0.44\textwidth]{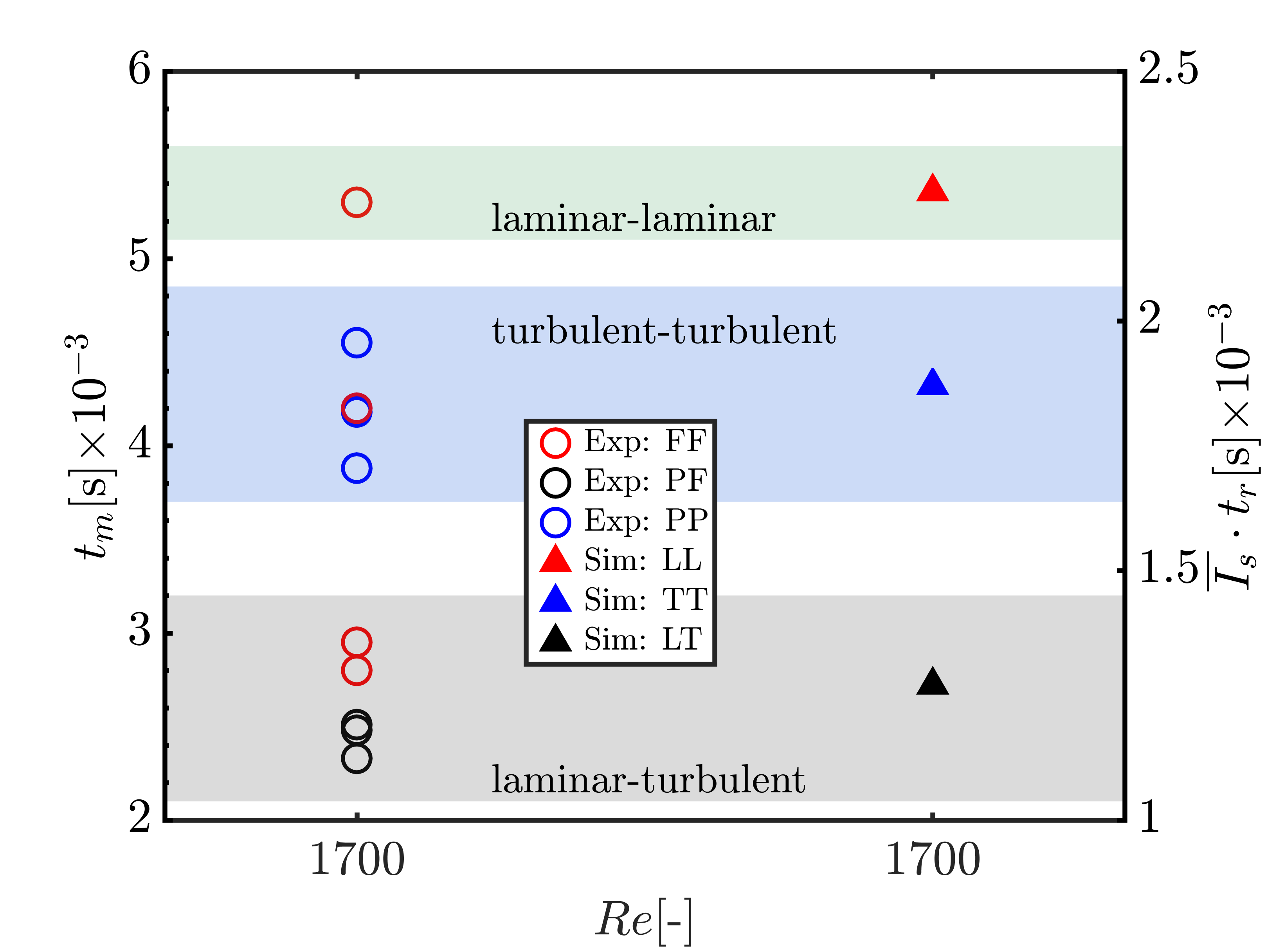}
		\caption{\label{fig:comp1700}Experimentally measured mixing time $t_m$ (triangles) and numerically computed intensity of segregation $I_s$ multiplied with the mean residence time $t_r$ (circles) at $Re=1700$. In both cases, three different inflow conditions were used: free-free (FF), perturbed-free (PF) and perturbed-perturbed (PP) in the experiments, and analogously laminar-laminar (LL), turbulent-turbulent (TT) and laminar-turbulent (LT) in the simulations. }		
	\end{center}
\end{figure}

At $Re\gtrsim 650$ the hysteresis region vanished and only the symmetric state remained regardless of the boundary conditions. To be more precise, the time-averaged velocity field flow is left-right symmetric, but the flow is turbulent, and vortices of different sizes enlarge the fluid interface locally (macromixing), thereby enhancing mixing by molecular diffusion (micromixing). As $Re$ was increased, the turbulent fluctuations raised in the collision zone and the Kolmogorov scale $\eta$ diminshed progressively. 

The error bars in Fig.~\ref{fig:micro_exp_tmix}(b) show that for $Re\lesssim 1400$ and $Re\gtrsim 2000$ there was little scatter in the experimentally measured $t_m$, whereas in between large deviations were observed.  The perfect overlap of the transitional regime in square-duct flow with the regime in which large scatter was observed in our experiments, strongly suggested that imperfections at the entrance of the lateral ducts triggered turbulent puffs, which then proceeded toward the junction of the T-mixer. We tested this hypothesis both in simulations and experiments.

In the numerical simulations, the flow at the inlet can be prescribed as turbulent or laminar, as explained in \S\ref{sec:num}. Although in our experiments there could be a random occurrence of turbulent puffs separated by laminar segments at $Re\in[1400,2000]$, the details of this process are setup-specific and very expensive to simulate\cite{avila2013}. For simplicity, we simulated three different scenarios at $Re=1700$: laminar flow at both inlets (LL), turbulent flow at both inlets at (TT) and a mixed case (LT), with turbulent flow at the left inlet ($y<0$) inlet and laminar flow at the right inlet ($y>0$). For the turbulent inlets, fully turbulent square duct was computed at $Re=4000$ and was downscaled by a factor of $4000/1700=2.35$ to generate the inlet boundary conditions for $Re=1700$. These three distinct inflow conditions can be seen as the limiting cases for the random collision of traveling puffs expected in experiments. 

The temporal evolution of the intensity of segregation $I_s$ with these inlet flow conditions is shown in Fig.~\ref{fig:dom1700}. For laminar inlets (LL), the average intensity of segregation was $\overline{I_s}=0.32$, whereas for turbulent inflows (TT) the mixing was better ($\overline{I_s}=0.27$). Surprisingly, the mixed case (LT) yielded a vast improvement of the mixing efficiency with $\overline{I_s}=0.18$. The reason for this behavior stems from the shape of the collision interface at the junction of the T-mixer, which initiates and determines the development of the mixing process along the outlet channel. Fig.~\ref{fig:tke_cross} shows snapshots of the turbulent kinetic energy $k$ at $x=0.25$ (top row) and $x=1.5$ (bottom row) in the mixing channel for the three inflow cases considered. For laminar inflows, the influence of the parabolic shape of the laminar inflow profile is seen in the collision interface, which yields the highest turbulent intensities at the top and bottom walls and a thin layer of quiescent intensities at the peak of the parabola, see Fig.~\ref{fig:tke_cross}(a). By contrast, for the turbulent inflows a broad interface with the highest intensities in the center was found, see Fig.~\ref{fig:tke_cross}(b). For the mixed inflow conditions, a biased parabolic shape with the highest intensities was found, but in contrast to LL case, the highest intensities are distributed along the parabola and not at the walls. This left-right asymmetry of the velocity field persists along the mixing channel (see the bottom row of Fig.~\ref{fig:tke_cross}) and accounts for the better mixing. While in the LL and TT cases the mixing is driven only by the fluctuations due to turbulent vortices, in the LT case the mean flow pattern also contributes to the mixing of the inlet streams, in an analogous manner to the asymmetric flow pattern of the engulfment regime. Note that in all cases the Reynolds number was identical at the two inlets, so the mass flow rate of the colliding streams was identical, and the impingment point remained in average at the center. 

In order to avoid the random generation of turbulent puffs in the experiments, we introduced helices made of wire in the cone of the inflow pipe (entrance to the square duct). This produced a turbulent inflow as used in the simulations. Three distinct inlet conditions were tested at $Re=1700$ (free-free: FF, perturbed-free: PF, perturbed-perturbed: PP). When the disturbance was applied at both inlets (PP), the micro-mixing time was $t_m \approx 4.3\times 10^{-3}$s and exhibited a small standard deviation ($2\times10^{-4}$s). When only one of the two inflows was disturbed (PF), mixing was greatly enhanced with $t_m\approx2.4\times 10^{-3}$s. We performed four experimental runs with two undisturbed inlets (FF), and observed a great variation depending on the experimental realization. In two cases, the same $t_m$ as PP was observed, in one instance as in PF, and in another instance, much poorer mixing was observed, exactly as expected from the collision of two laminar inflows. The comparison of the experimentally measured $t_m$ with the numerically computed $I_s$ is shown in Fig.~\ref{fig:comp1700} and confirms the influence of the inflow condition on the quality of mixing. 

\subsection{Fully turbulent regime}\label{sec:fullturbulent}
\begin{figure}[h]
	\begin{center}
			(a)\\
\includegraphics[width=0.41\textwidth]{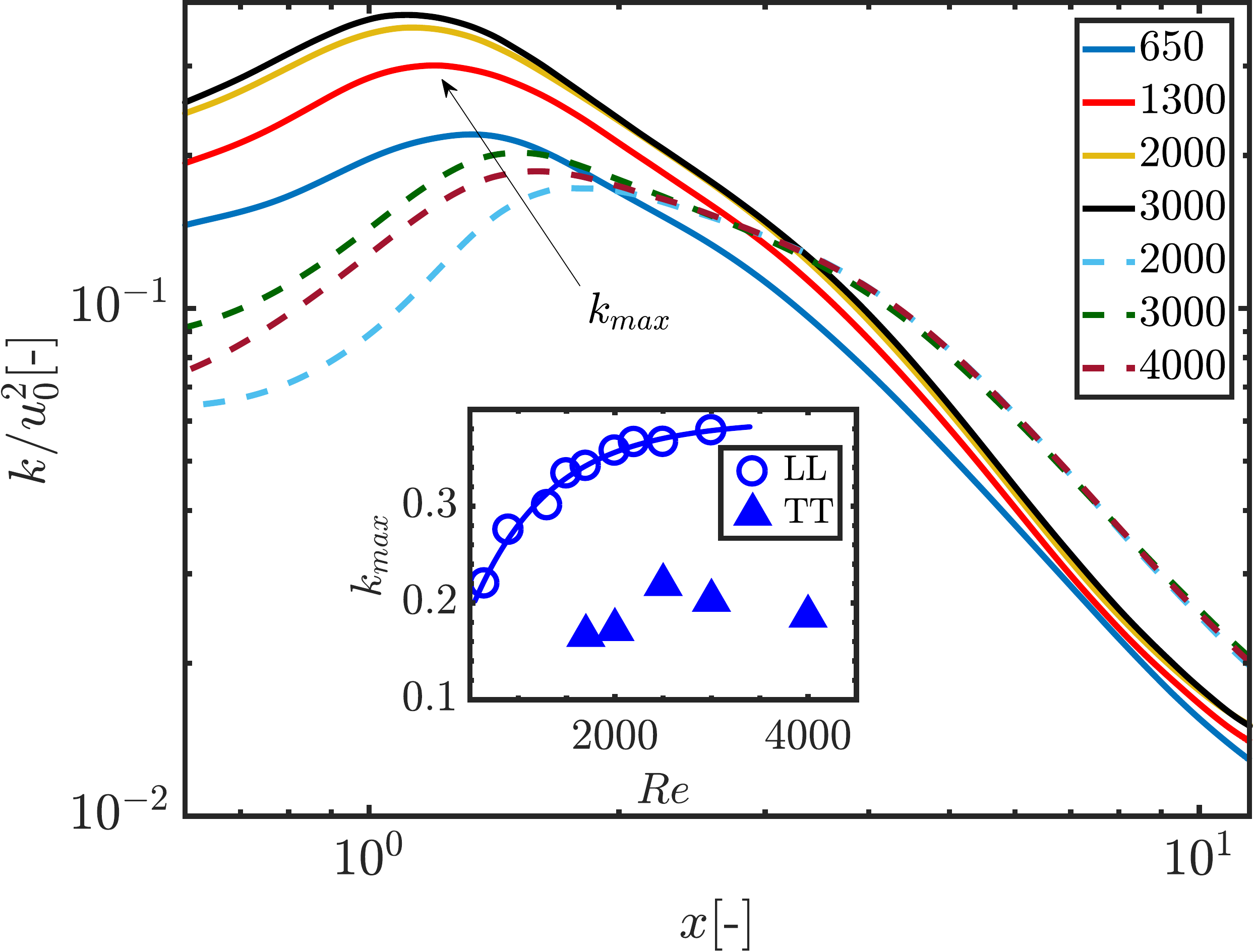}
\\(b)\\
\includegraphics[width=0.41\textwidth]{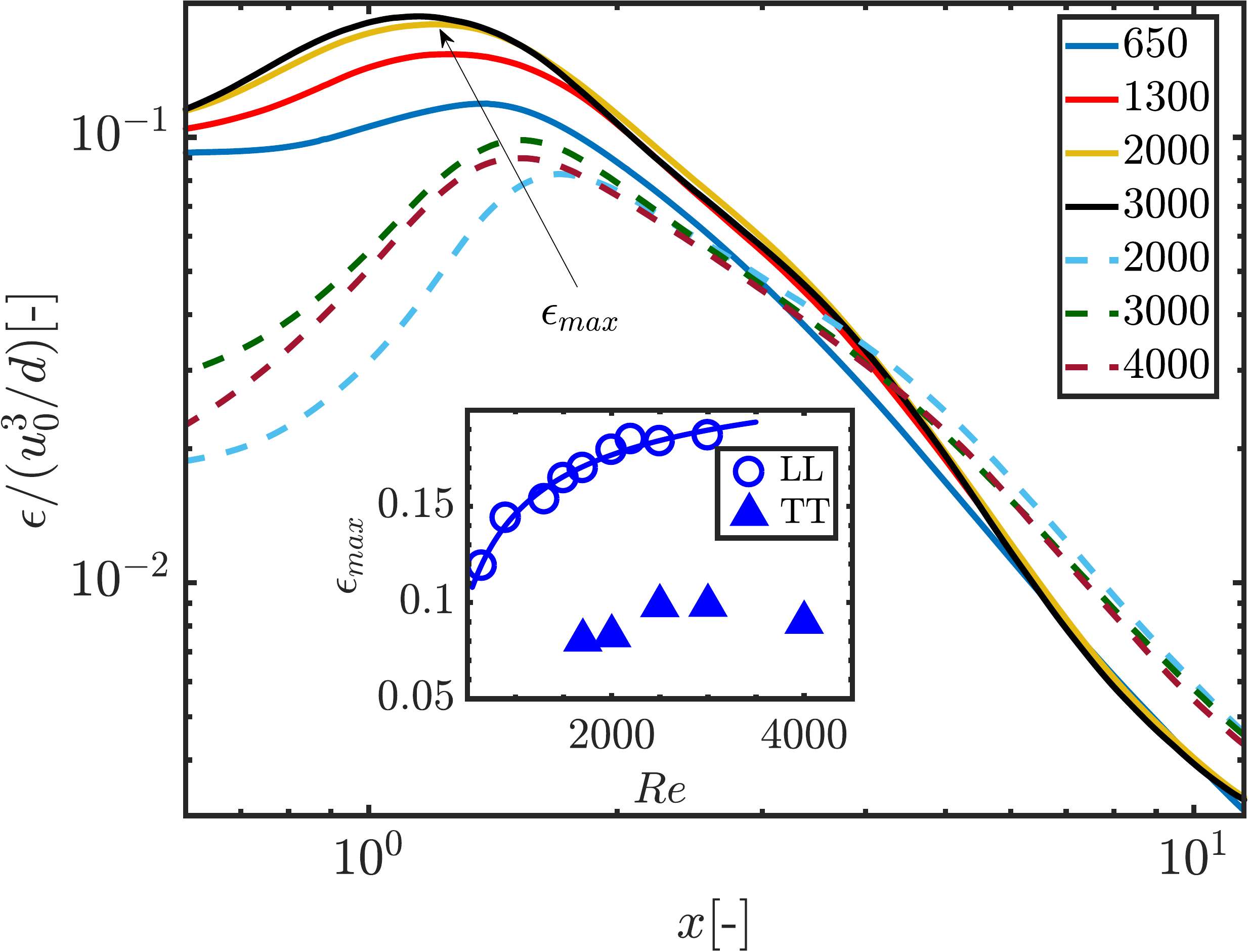}	
		\caption{(a) Normalized turbulent kinetic energy $k/u_0^2$ and dissipation (b) $\epsilon/(u_0^3/d)$ along the mixing channel for various Reynolds numbers with laminar (solid line) and turbulent (dashed line) inflow conditions. The inset shows the maximum value $k_\text{max}$ and $\epsilon_\text{max}$ of $k/u_0^2$ and $\epsilon(u_0^3/d)$, respectively, as function of $Re$.}
		\label{fig:kundeps_pos}
	\end{center}
\end{figure}
As the Reynolds number was increased beyond $Re>2000$, the mixing time $t_m$ continued to improve, but much more slowly as shown in~Fig.~\ref{fig:mixing_regime}. While such an asymptotic behavior of $t_m$ was observed experimentally in several  mixers \cite{nagasawa2005design,aoki2006effects,kashid2011mixing,zhendong2012mixing,kockmann2006convective,siddiqui2009characteristics}, it remains poorly understood. Our simulations resolve all temporal and spatial scales in turbulence, allowing a more detailed analysis than possible in past studies. Figs.~\ref{fig:kundeps_pos} shows the evolution of the normalized turbulent kinetic energy $k/u_0^2$ and dissipation $\epsilon/(u_0^3/d)$, both averaged over the cross-section, along the mixing channel axis for several $Re$ in the turbulent regime. Initially, both quantities increase as turbulence intensifies due to the collision of the two incoming streams, and subsequently, a maximum is reached before turbulent fluctuations (and thus dissipation) begin to decay. This onset of decay also marks the onset of strong mixing, and the decay behavior is fairly independent of the Reynolds number $Re$ and is mainly determined by the inflow conditions (LL/TT). The insets of Figs.~\ref{fig:kundeps_pos} show that for each inflow condition, the maximum values of the normalized turbulent kinetic energy and dissipation increase, but finally saturate as $Re$ increases. This is in agreement with the asymptotic theoretical scalings $k\propto Re^2$ and  $\epsilon \propto Re^3$ for isotropic turbulence. As turbulence in the T-mixer reaches the asymptotically behavior, the improvement of the mixing efficiency also approaches an asymptotic state and results in a slow decrease of $t_m$, as observed for $Re\gtrsim 2000$. 

Note also that the LL case exhibits larger dissipation than TT, which may be surprising in view that turbulent inlets dissipate more than  laminar ones. However,  laminar inflow profiles are quasi-parabolic and thus carry much more kinetic energy than flat turbulent inflow profiles (see Fig.~\ref{fig:inflow_cross}). This excess energy results in larger dissipation in the main (mixing) channel. As the dissipation at the inlets is negligible compared to the dissipation in the mixing channel, laminar inlets cause a larger net dissipation.

\section{Specific power input and mixing time}
\label{sec:mixingmodels}

\begin{figure}[h]
	\begin{center}
(a)\\
\includegraphics[width=0.44\textwidth]{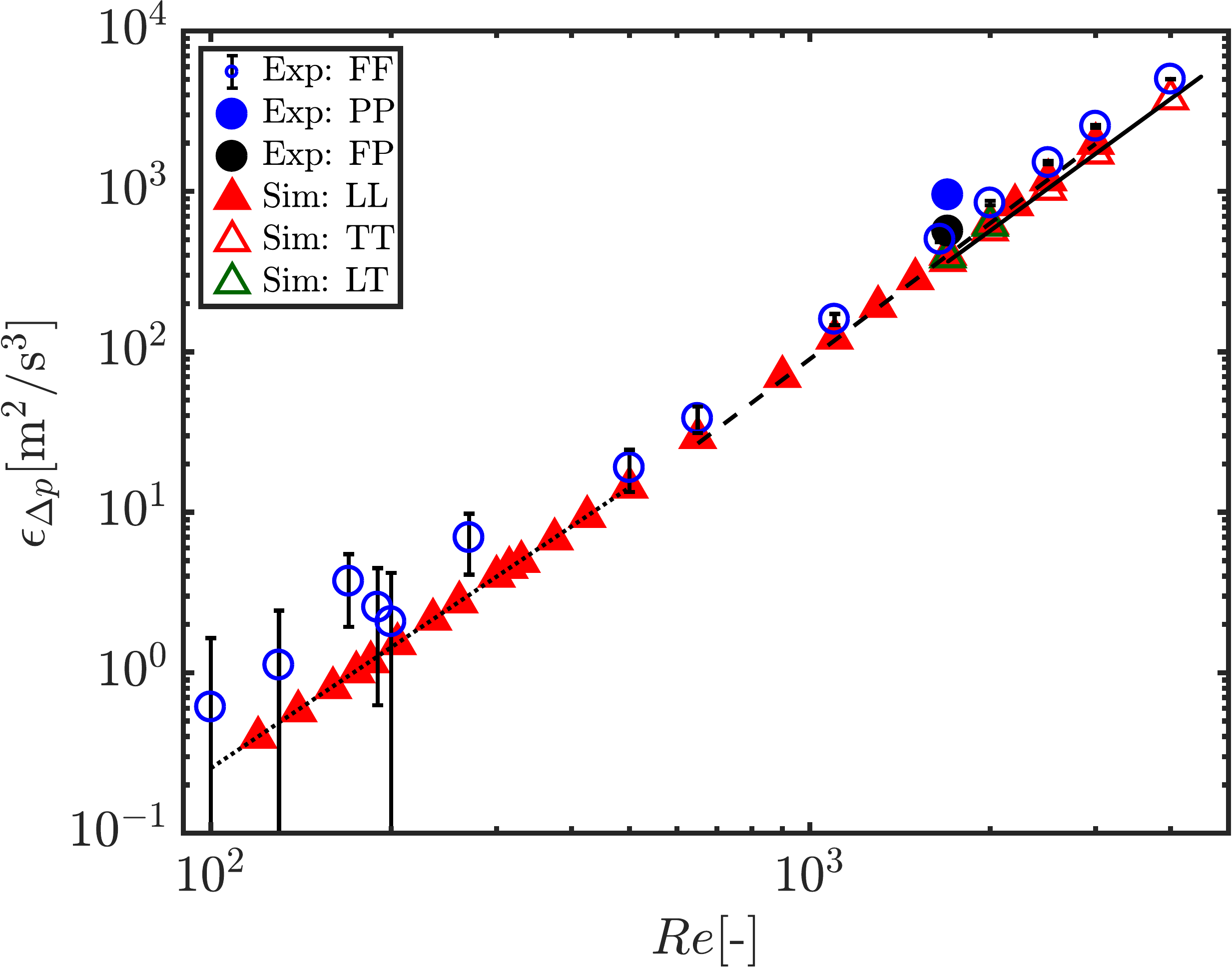} \\
(b)\\
\includegraphics[width=0.44\textwidth]{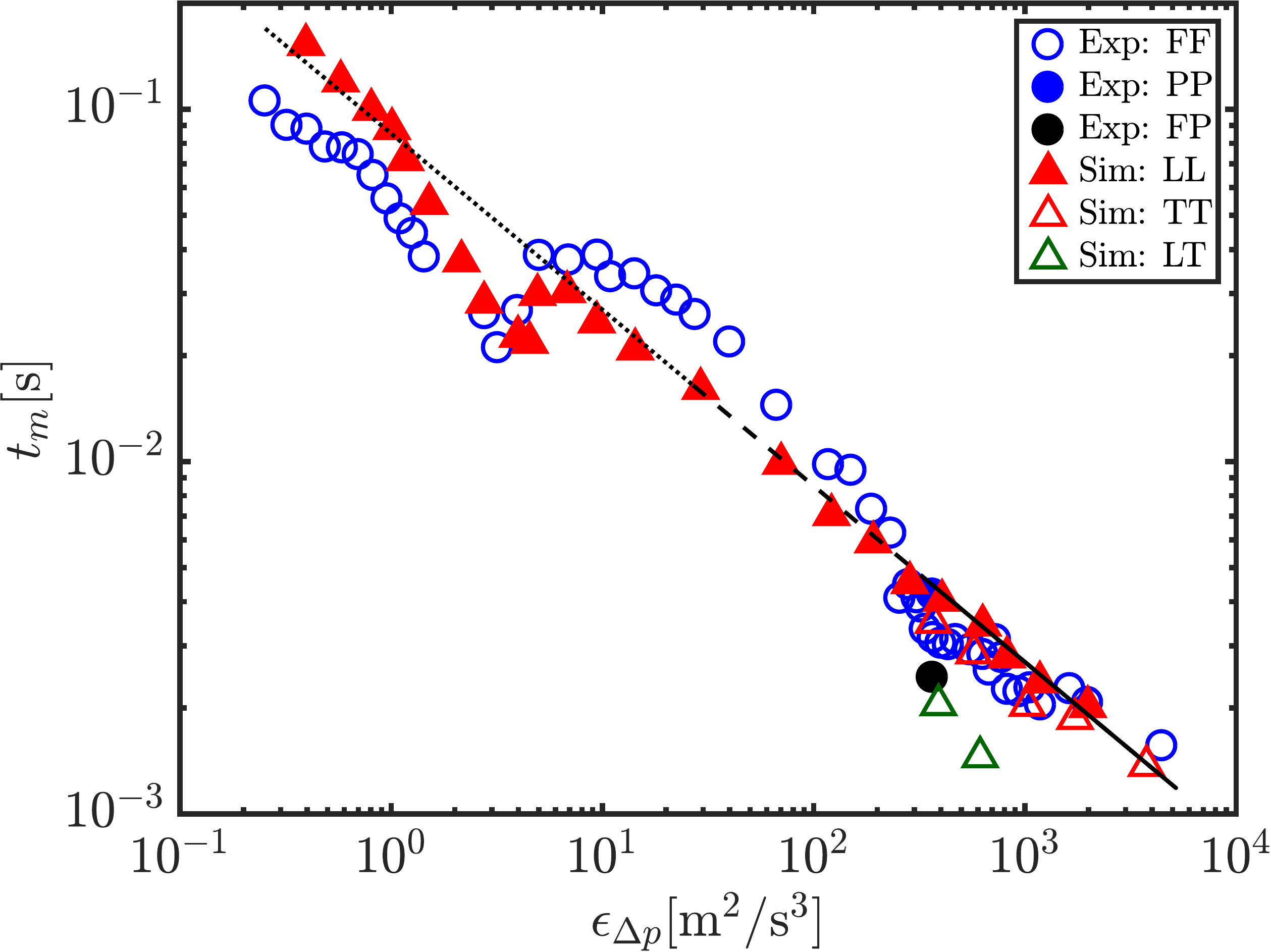}
		\caption{\label{fig:dissipationDP}(a) Turbulent dissipation $\epsilon_{\Delta p}$ estimated from the pressure loss $\Delta p$ as function of Reynolds number $Re$ comparing experiments and simulations while considering all three inflow scenarios: free-free (FF), perturbed-free (PF) and perturbed-perturbed (PP) in the experiments, and analogously laminar-laminar (LL), turbulent-turbulent (TT) and laminar-turbulent (LT) in the simulations. (b) Experimentally measured mixing time $t_m$ and computationally estimated mixing time using Eq.~\ref{eq:equality} as a function of energy dissipation $\epsilon_{\Delta p}$ in $Re=(90,4000)$. In case of $t_m$, $\epsilon_{\Delta p}$ is estimated from the power law fit obtained in (a). Straight lines show the theoretical relation $t_m=(\nu/\epsilon_{\Delta p})^{0.5}$.}
	\end{center}
\end{figure}

Mixing models such as the interplay of exchange with the mean (IEM) \cite{commenge2011villermaux}, the engulfment diffusion deformation (EDD) \cite{baldyga1999turbulent} or the incorporation \cite{villermaux1994generalized} micromixing model, are widely used to predict and characterize mixing and reaction outcomes. They especially find application at high Schmidt number flows to quantify micromixing. Their common underlying assumption is that within the lifetime of a small eddy $t_\eta = \left(\nu/\epsilon\right)^{0.5}$ micromixing is completed, thus assuming implicitly that the turbulent dissipation determines mixing. In practice, the turbulent dissipation is typically estimated as $\epsilon_{\Delta p} = \Delta p \,Q/(\rho V)$, where $\Delta p$ is the pressure loss, $Q$ is the volume flow rate, and $V$ is the volume of fluid in the T-mixer. The micromixing time is then estimated in the aforementioned models as $t_m\propto t_\eta\simeq(\nu/\epsilon_{\Delta p})^{0.5}$, see \citet{falk2010performance}. 

Fig.~\ref{fig:dissipationDP}(a) shows that in our T-mixer $\epsilon_{\Delta p}$ is nearly independent of the inlet boundary conditions, whereas the mixing time is mainly determined by them. In fact, at the same $Re$ laminar inlets feature slightly higher dissipation than turbulent inlets, whereas the mixing time is lower for the latter. Remarkably, by using a laminar and a turbulent inlet at $Re=2000$ in the simulations, the same mixing time as $Re=4000$ with two turbulent inlets could be achieved. Given that the pressure loss is not much affected by the inlet conditions, this implies that the energy dissipation is about six times lower, as shown in Fig.~\ref{fig:dissipationDP}(a). In other words, the same mixing time can be achieved at $Re=2000$, but with a specific power input of only $\approx600$W/kg instead of the $\approx3500$W/kg needed to operate at $Re=4000$. Unfortunately, in our experiments we could only keep the inlets laminar up to $Re=1700$. By disturbing one inlet and keeping the other laminar, we could achieve the same mixing time as for $Re=2200$ with turbulent inlets. The energy dissipation was $\epsilon_{\Delta p}=560$W/kg and $1200$W/kg, respectively, \ie a factor of 2 lower. This demonstrates that the dissipated energy or input power can be a poor estimator of mixing, in contrast to the commonly accepted view and the underlying assumption of the aforementioned models, \ie $t_m\propto (\nu/\epsilon_{\Delta p})^{0.5}$. While our experimental and numerical data exhibit this scaling in the fully turbulent regime, the boundary conditions have a significant impact.

Finally, we stress that the dependence of the mixing time on the energy dissipation shown in Fig.~\ref{fig:dissipationDP}(b) is close to that obtained by \citet{falk2010performance} using a variety of mixers (see their Fig.~5). Our study covers the whole range of their mixing times, but for the same dissipation, our mixing times are roughly three times lower in average than theirs. For the specific case of mixed laminar-turbulent inflow conditions at $Re=2000$, our mixing time is even six times lower than theirs.

\section{Conclusion}

Turbulence models based on the Reynolds-averaged Navier--Stokes equations (RANS) or simple empirical micromixing models are typically used to design chemical reactors. Their accuracy, and more importantly, their transferability to new situations is however doubtful. Because of the increase in computing power in the last decades years, DNS of fluid flows at operationally relevant regimes are now possible. They allow accurate predictions of turbulent mixing and comparisons to laboratory experiments, as done here for a simple T-shaped mixer. The highest attained $Re=4000$ is well beyond previous experimental and numerical works, which have mostly focused on moderate Reynolds numbers, \eg up to $Re=400$ \cite{mariotti2018steady} in a T-mixer and $Re=1100$ in Confined Impinging Jet Reactor \cite{gradl2009simultaneous,schwertfirm2007flow}.

A detailed macroscopic description of mixing as presented here is key to predict the qualitative outcome of mixing-sensitive chemical reactions or precipitation processes in the liquid phase. The good agreement between the numerically computed segregation index $I_s$ and experimentally measured mixing time $t_m$ support the validity of our approach. This suggests a predominant role of turbulent eddies (or macromixing) in controlling the mixing efficiency. However, caution must be taken because of the freedom in the factor used to compare the two data sets, \ie here $t_m=1.8\,I_s\cdot t_r$. Changing this factor allows an arbitrary vertical shift between the two data sets in Fig.~\ref{fig:mixing_regime}. Despite this fact, the agreement in the trends (and slopes of the curves) indicates that either our numerical scheme models micromixing in a reasonable fashion, or that micromixing can be easily parametrized if macromixing is well resolved in the simulations. This result is very encouraging and suggests that by solving for chemical reactions simultaneously with the Navier--Stokes equations should lead to a quantitative prediction of mixing-sensitive reactions. 

Finally, our experimental and numerical findings reveal the importance of inlet boundary conditions in determining the efficiency of mixing. More specifically, we could show that by using mixed laminar-turbulent inflow conditions a six-fold reduction of the specific power input could be reached at constant mixing time. Future work will include experiments with longer inlet channels and precisely controlled inflow conditions to test the limits of this technique. For example, in continuous chemical synthesis, fast mixing is key for mixing-sensitive reactions. However, because of the exponential raise in the viscosity of the solvent with temperature (operating with external cooling), the increase of the pressure drop, thus the inflow rate, is limited in practice\cite{kim2016submillisecond}. Manipulating  the boundary conditions might be a promising and simple method to enhance mixing at low Reynolds numbers as well. 

\section*{Conflicts of interest}

There are no conflicts to declare.

\section*{Acknowledgements}
The authors would like to acknowledge the funding of the Deutsche Forschungsgemeinschaft (DFG) through the Cluster of Excellence Engineering of Advanced Materials (EAM) and Bayer Technology Services GmbH (BTS). The authors gratefully acknowledge the compute resources and support provided by the Erlangen Regional Computing Center (RRZE).


\balance

\renewcommand\refname{References}
\bibliography{biblio_Holger} 
\bibliographystyle{rsc} 

\end{document}


{\LARGE
\begin{center}
\bf{Supplementary information}
\\
\vspace{1cm}
\Large
\bf{Inflow boundary conditions determine T-mixer efficiency}
\\
\vspace{0.2cm}
\normalsize
\normalfont
\noindent\large{Tobias Schikarski,\textit{$^{a}$} Holger Trzenschiok,\textit{$^{a}$} Wolfgang Peukert,\textit{$^{a}$} and Marc Avila\textit{$^{b}$}}
\end{center}
}

\textit{$^{a}$~Institute of Particle Technology, Department of Chemical and Biological Engineering, Friedrich-Alexander-Universit\"at Erlangen-N\"urnberg, 91058 Erlangen, Germany; E-mail: \textcolor{blue}{\underline{tobias.schikarski@fau.de}}}\\
\textit{$^{b}$~Center of Applied Space Technology and Microgravity, Universit\"at Bremen, Am Fallturm, 28359 Bremen, Germany }





\tableofcontents
\section{Transition to turbulence}
\begin{figure}[!htbp]
\begin{center}
\includegraphics[width=0.5\textwidth]{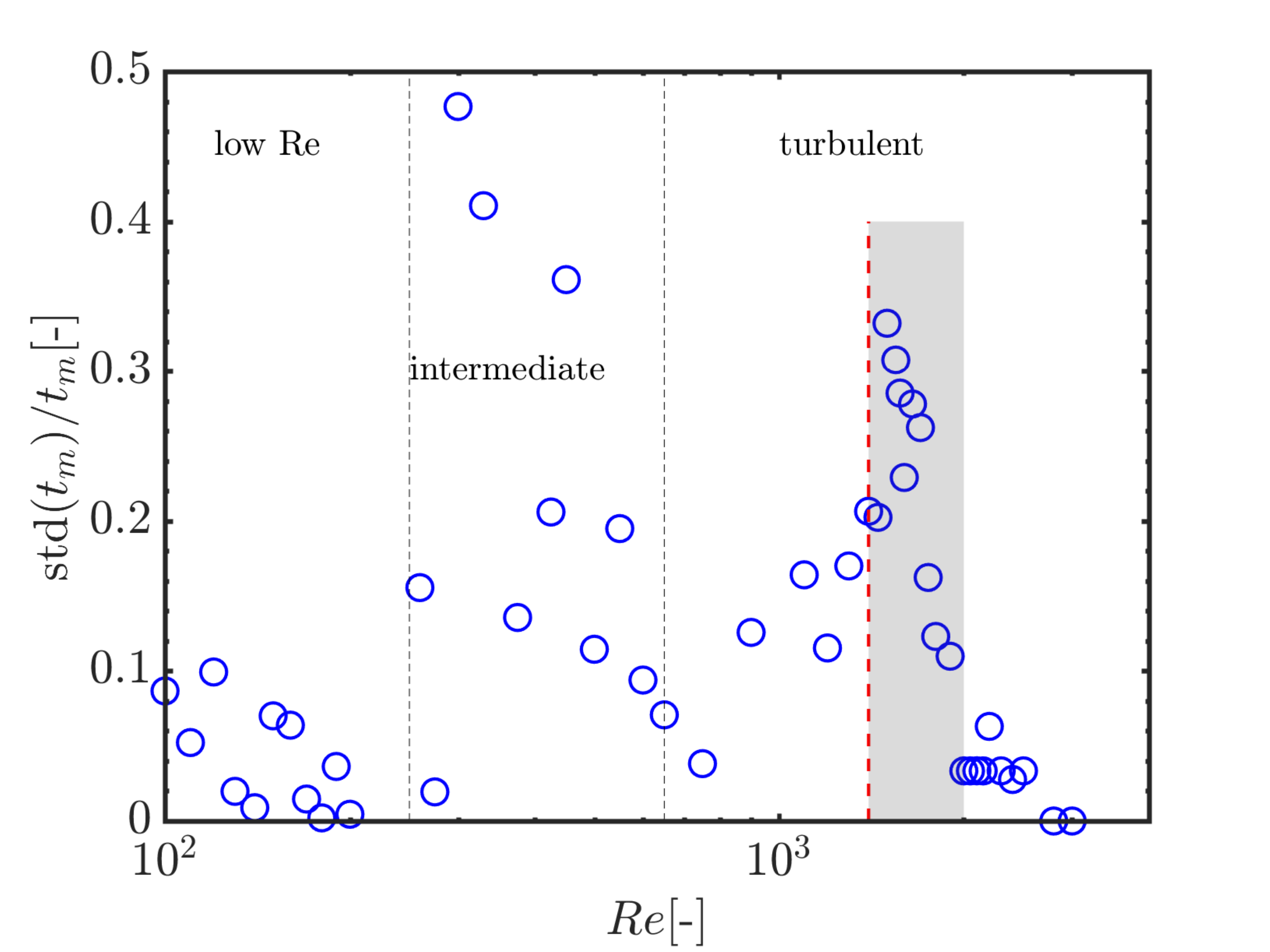}
\caption{\label{fig:std_mixingtime} Standard deviation of the experimentally obtained mixing time, scaled with the mean $\text{std}(t_m)/t_m$, as function of Reynolds number $Re$. The gray-shaded area depicts the Reynolds number range where turbulent puffs are expected to occur, whereas the red dashed line marks the natural onset of transition, which in our experiments is at $Re\approx1400$.}
\end{center}
\end{figure}
In section \S3, we argue that the natural and random occurrence of turbulent puffs beyond $Re\gtrsim 1460$ influences mixing in the outlet channel and results in a large deviation in mixing time $t_m$ (see Fig.~1b in paper). To support this statement, the standard deviation of mixing time, normalized by the mean $\text{std}(t_m)/t_m$, is shown as a function of Reynolds number in Fig.~\ref{fig:std_mixingtime}. Clearly, the deviation in measured mixing time increases strongly in the regime where turbulent puffs can occur. Besides, large deviations are also observed in the intermediate regime, where engulfing and symmetric flow structures compete. This emphasizes the general sensitivity of mixing on the inflow in the T-mixer.

\section{The intensity of turbulence in a T-mixer}

The collision of inlet streams is used in common mixing devices to trigger strong turbulent motions, hence mixing, at relatively low inflow Reynolds numbers. The accurate computation of the turbulent statistics in our simulation method allows to estimate the turbulent intensity in the mixing channel and compare it to canonical flows, \eg pipe flow. To do so, the definition of the Taylor microscale  for homogeneous isotropic turbulence, $\lambda= \sqrt{10\nu k/\epsilon}$, where $k$ and $\epsilon$ are the turbulent kinetic energy and the dissipation, respectively, is assumed to hold. 
Then, the turbulent Reynolds number $Re_\lambda=\sqrt{k}\lambda/\nu$ can be calculated by extracting the values of $k$ and $\epsilon$ from Figure 10. Taking the maximum values of $k$ and $\epsilon$ (at around $x=(1,2)$), we obtain $Re_\lambda\approx50$ and $Re_\lambda\approx125$ for $Re=650$ and $Re=4000$, respectively. Comparing these values to pipe flow (S.C.C. Bailey \textit{et al.}\cite{bailey2009measurement} we are not aware of any study in square duct flow providing similar data), the turbulence level in a T-mixer at $Re=4000$ is comparable to $Re\approx60000$ in cylindrical pipe flow. 

\begin{figure}[!htbp]
\begin{center}
\includegraphics[width=0.5\textwidth]{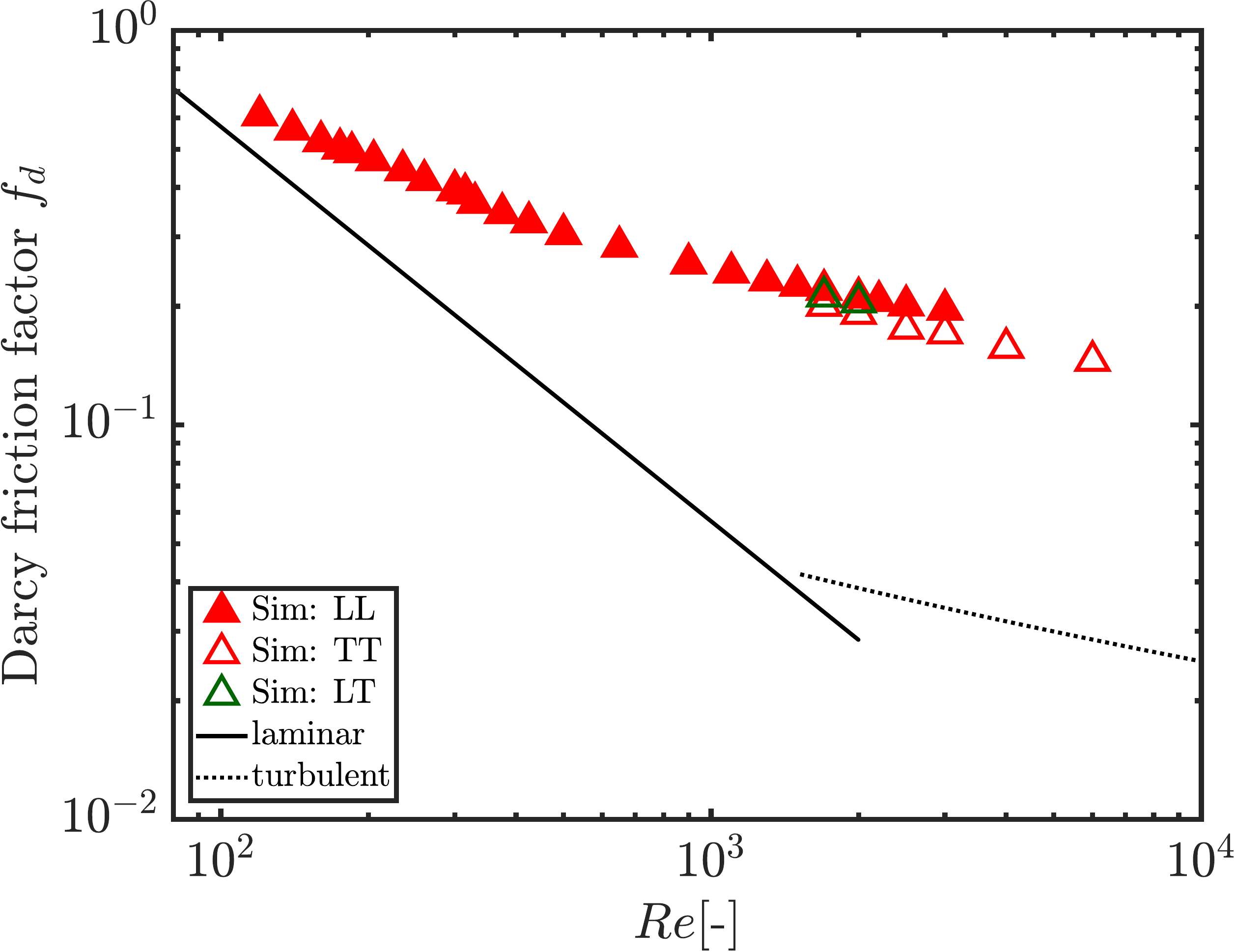}
\caption{\label{fig:darcy} Darcy friction factor, as defined in Eq.~\eqref{eq:darcy},  as function of Reynolds number $Re$. Symbols represent the data obtained from simulation in the T-mixer. The solid and the dashed line denote the laminar and turbulent friction factor of square duct-flow, see Eqs.~\eqref{eq:darcy_l} and \eqref{eq:darcy_t}  respectively. }
\end{center}
\end{figure}

A more direct comparison to square-duct flow can be done by comparing the pressure loss $\Delta p$ of our T-mixer to that of a square duct of the same length by using the Darcy friction factor 
\begin{equation}\label{eq:darcy}
f_d = 2\frac{\Delta p d_h}{L \rho u^2_0},
\end{equation}
where $d_h$ is hydraulic diameter (of the outlet in case of the T-mixer), $u_0$ is the mean velocity and $L$, here 15.5$d$, is the length between inlet and outlet. In square duct flow, the Darcy friction factor is 
\begin{equation}\label{eq:darcy_l}
f_d=\dfrac{57}{Re},
\end{equation}
for fully developed laminar flow, whereas in the fully turbulent regime it is commonly estimated with an empirical correlation proposed by Jones\cite{jones1976improvement}
\begin{equation}\label{eq:darcy_t}
f_d^{-1/2}=2\log_{10}(2.25Re f^{1/2})-0.8. 
\end{equation}
Fig.~\ref{fig:darcy} shows that at low $Re$ the friction factor of the T-mixer is close the laminar value of square-duct flow. However, as $Re$ increases, the friction factor gradually departs from it because of the subsequent hydrodynamic instabilities, which result in turbulent flow at the junction. As the turbulence develops fully, the friction factor of the T-mixer appears to evolve nearly parallel to that of square-duct flow, but it is about 5 times larger in magnitude.
\section{Turbulent statistics of square duct flow}
\label{sec:App_TurbStat}
\begin{figure}[!htbp]
	\begin{center}
		\begin{tabular}{cc}
			(a) & (b) \\
\includegraphics[width=0.45\textwidth]{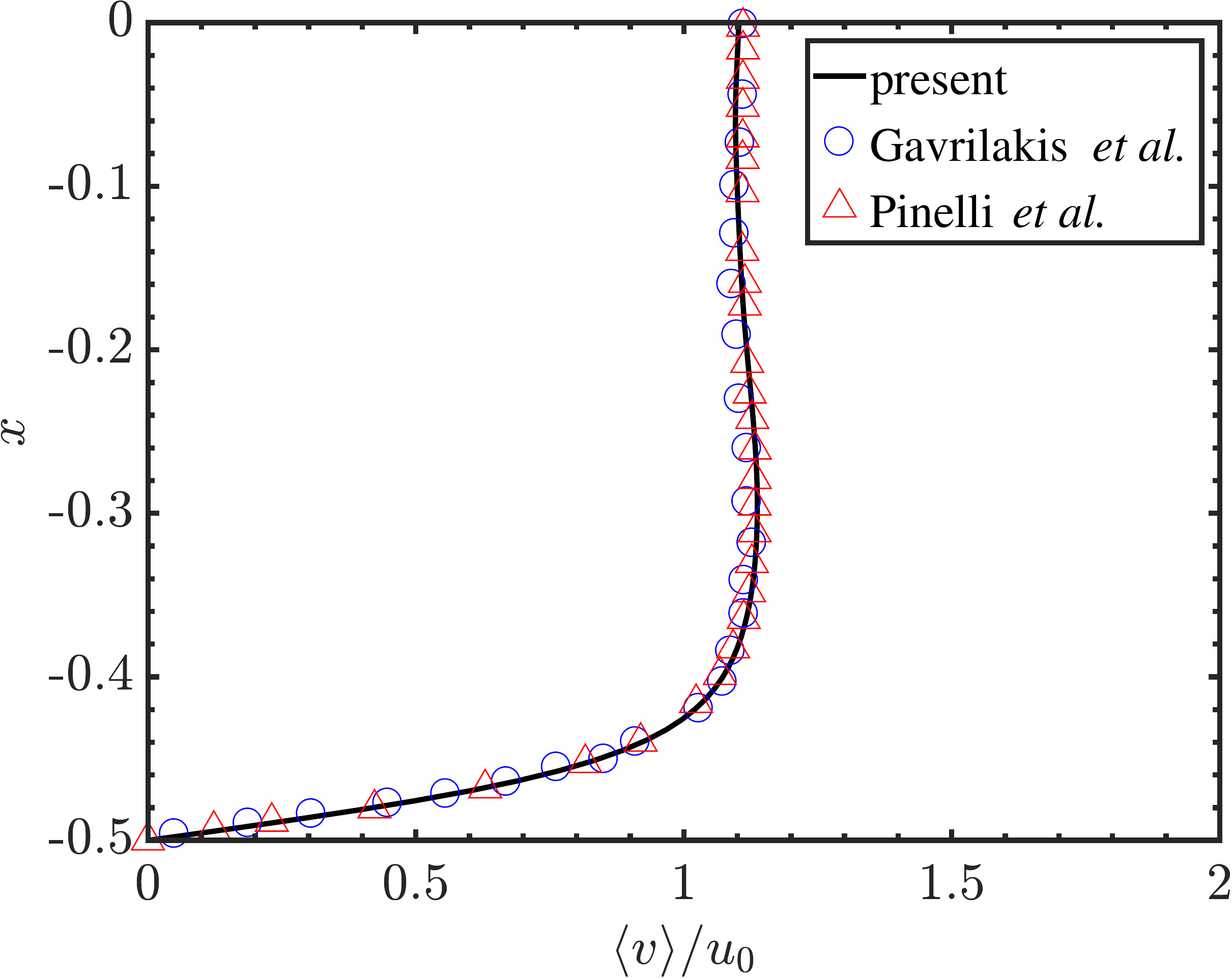}&
\includegraphics[width=0.45\textwidth]{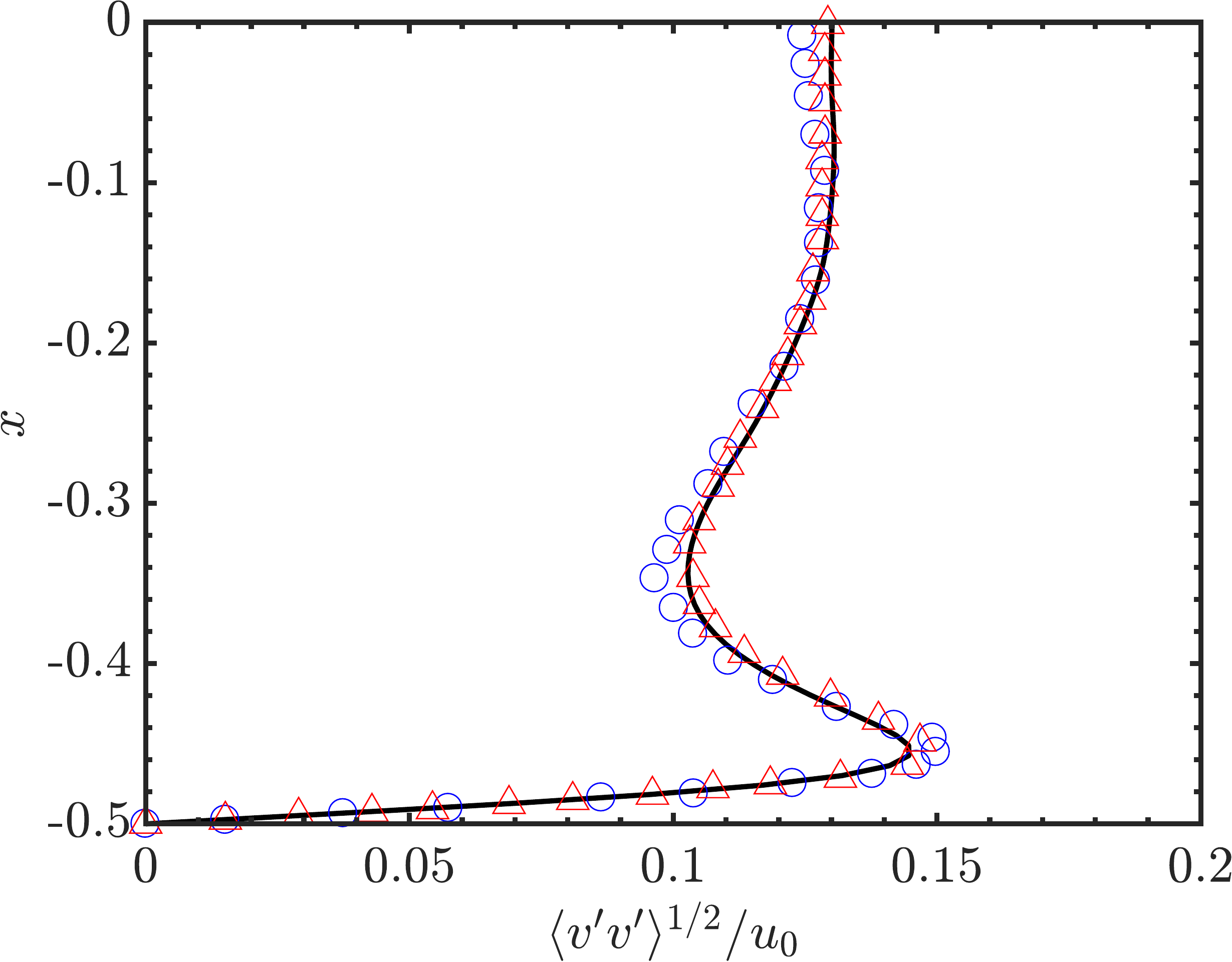}\\
(c) & (d)\\
\includegraphics[width=0.45\textwidth]{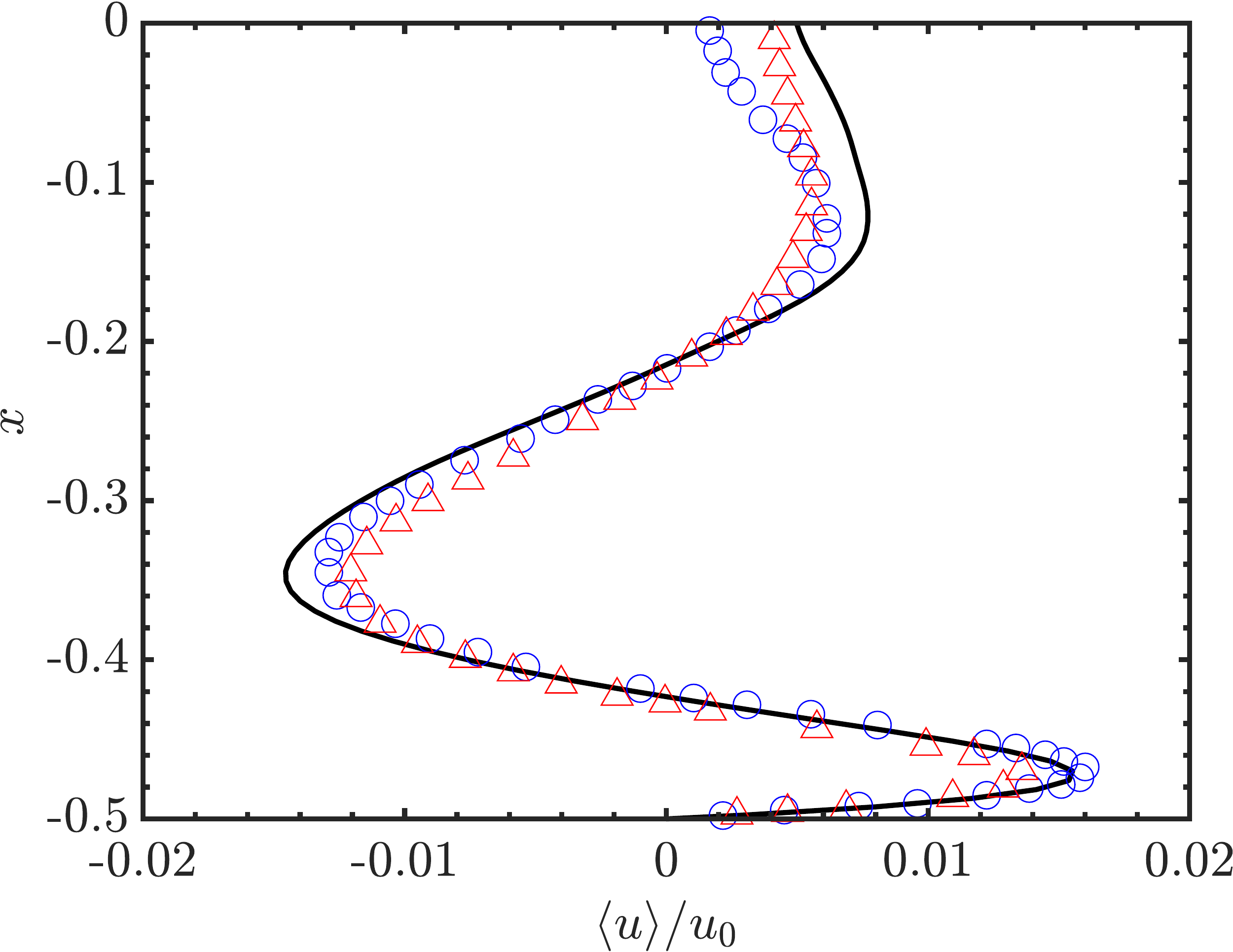}&
\includegraphics[width=0.45\textwidth]{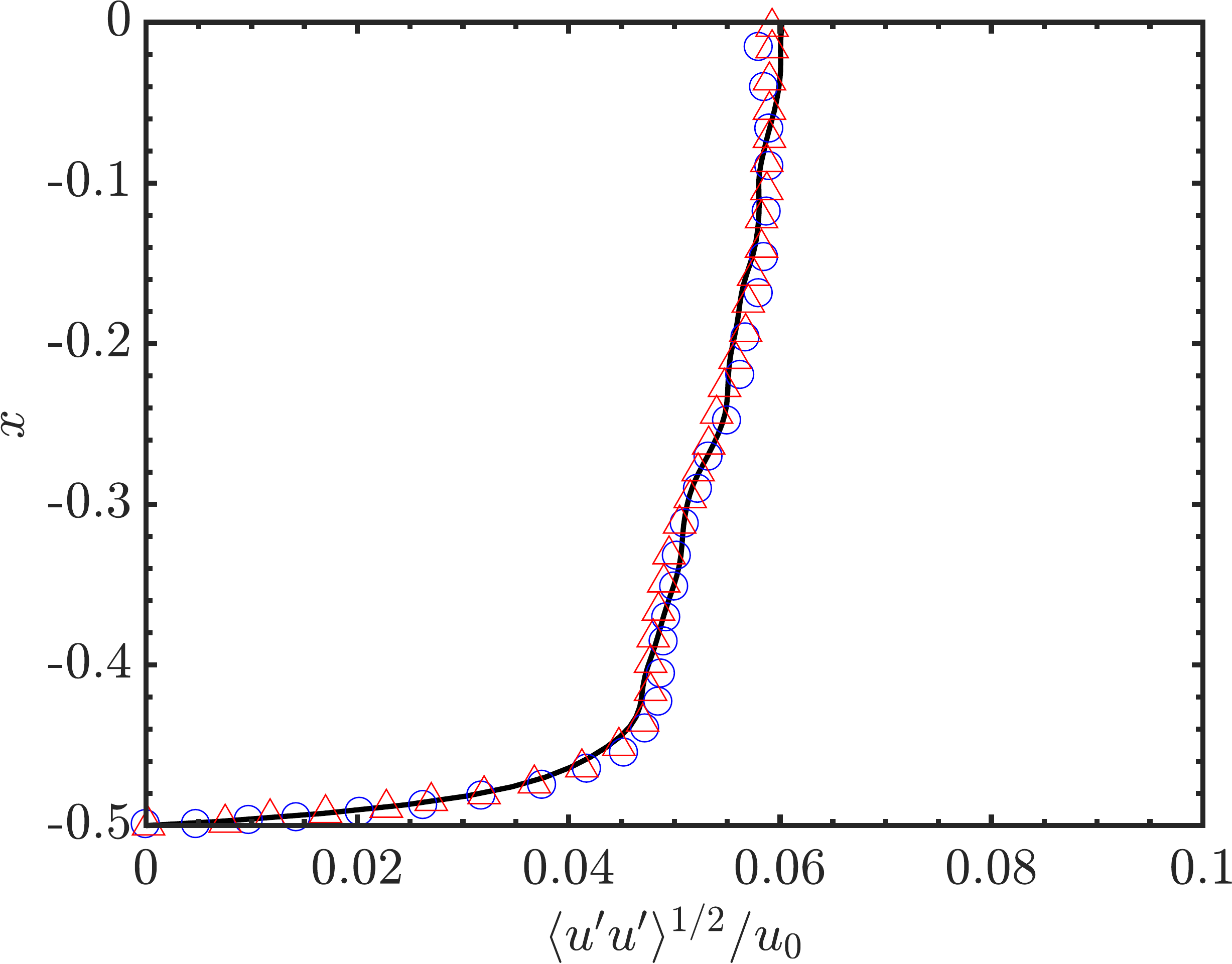}
		\end{tabular}
		\caption{\label{fig:turbstat}Turbulent statistics (black line) obtained at $Re=4000$ are compared to reference data (blue circles \cite{pinelli2010reynolds} and red triangles \cite{gavrilakis1992numerical}) obtained at $Re=4410$. (a) and (b) show the mean $\langle(\bullet)\rangle$ and root mean square velocity $\langle (\bullet)^\prime (\bullet)^\prime \rangle^{1/2}$ components at position $z=0.35$, respectively. The top and bottom row depict the streamwise $v$ and cross-stream $u$ velocity component, respectively. Note that the orientation of coordinates of the T-mixer inlet are used as well as only the half of the channel width is shown.}
		
	\end{center}
\end{figure}

In our simulations of the flow in a T-mixer at large $Re$, we apply fully developed turbulent flow at the inlet boundaries. To generate a fully developed turbulent inflow, we simulated the fluid flow at the desired Reynolds number in a periodic square duct with stream-wise length of $L=30d$ (see Fig.~6 in paper) and mapped the stored data for two selected cross sections to the two inlet boundaries of the T-mixer. To avoid correlated results, the data of the two cross sections is used with $100d/u_0$ delay. 

Fig.~\ref{fig:turbstat} shows a comparison of the first and second order turbulent statistics obtained with the our DNS to reference data\cite{pinelli2010reynolds, gavrilakis1992numerical} at a fixed height, here $z=0.35$. We averaged our data in time and in the (homogeneous) streamwise direction, which is denoted by the operator $\langle \rangle$. Good agreement between our simulations and the reference simulations in the mean velocity components $\langle u \rangle$ and $\langle v \rangle$ (see Fig.~\ref{fig:turbstat}(a)) was obtained. Slight deviations apart from the wall in cross-stream velocity component $\langle v \rangle$ are attributed to the too short average time interval. Similar, the normal Reynolds stresses  $\langle u^\prime u^\prime \rangle^{1/2}$ and $\langle v^\prime v^\prime \rangle^{1/2}$, where $u^\prime =u-\langle u \rangle$ describes the velocity fluctuations, agree very well to the reference data (see Fig.\ref{fig:turbstat}(b)).
\section{Pressure drop measurements}
\label{sec:pressdrop}
\begin{figure}[!htbp]
	\begin{center}
\includegraphics[width=0.4\textwidth]{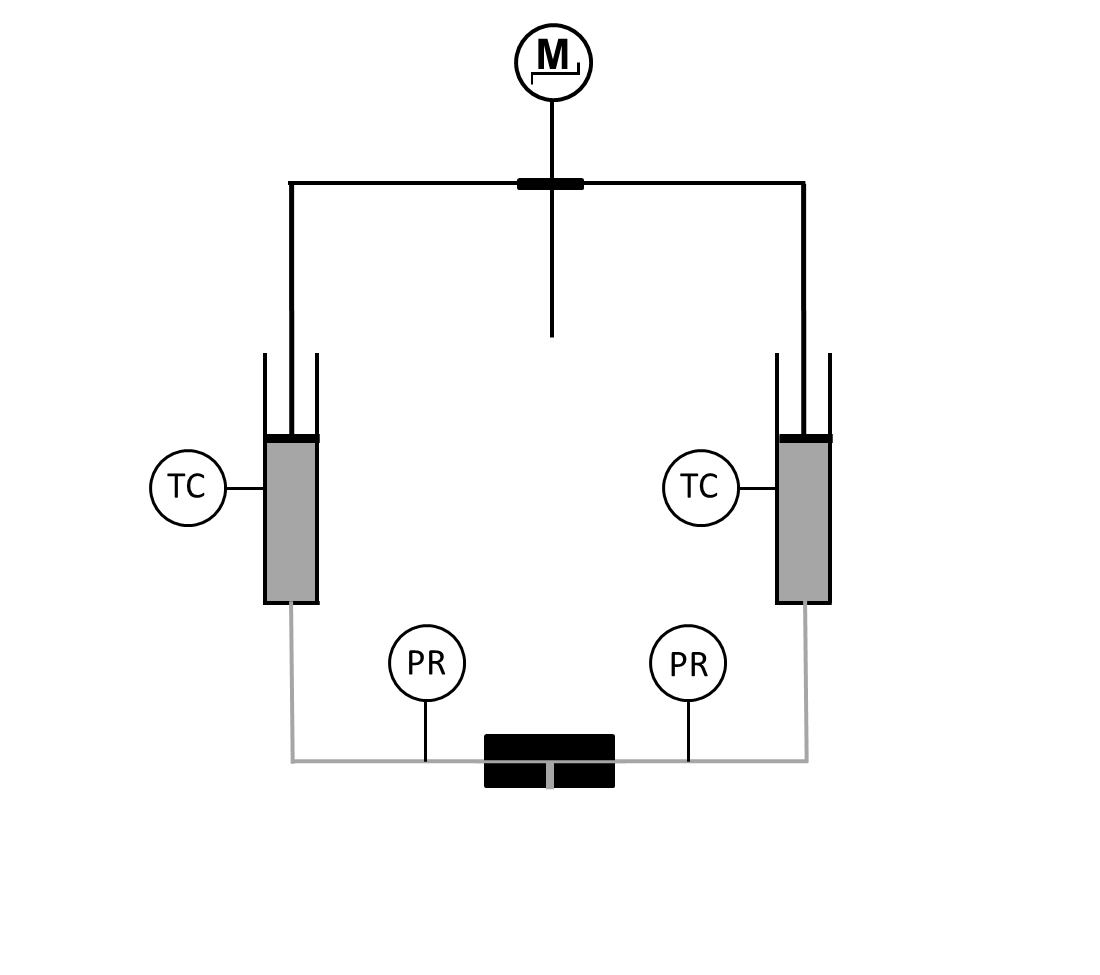}\\
\includegraphics[width=0.4\textwidth]{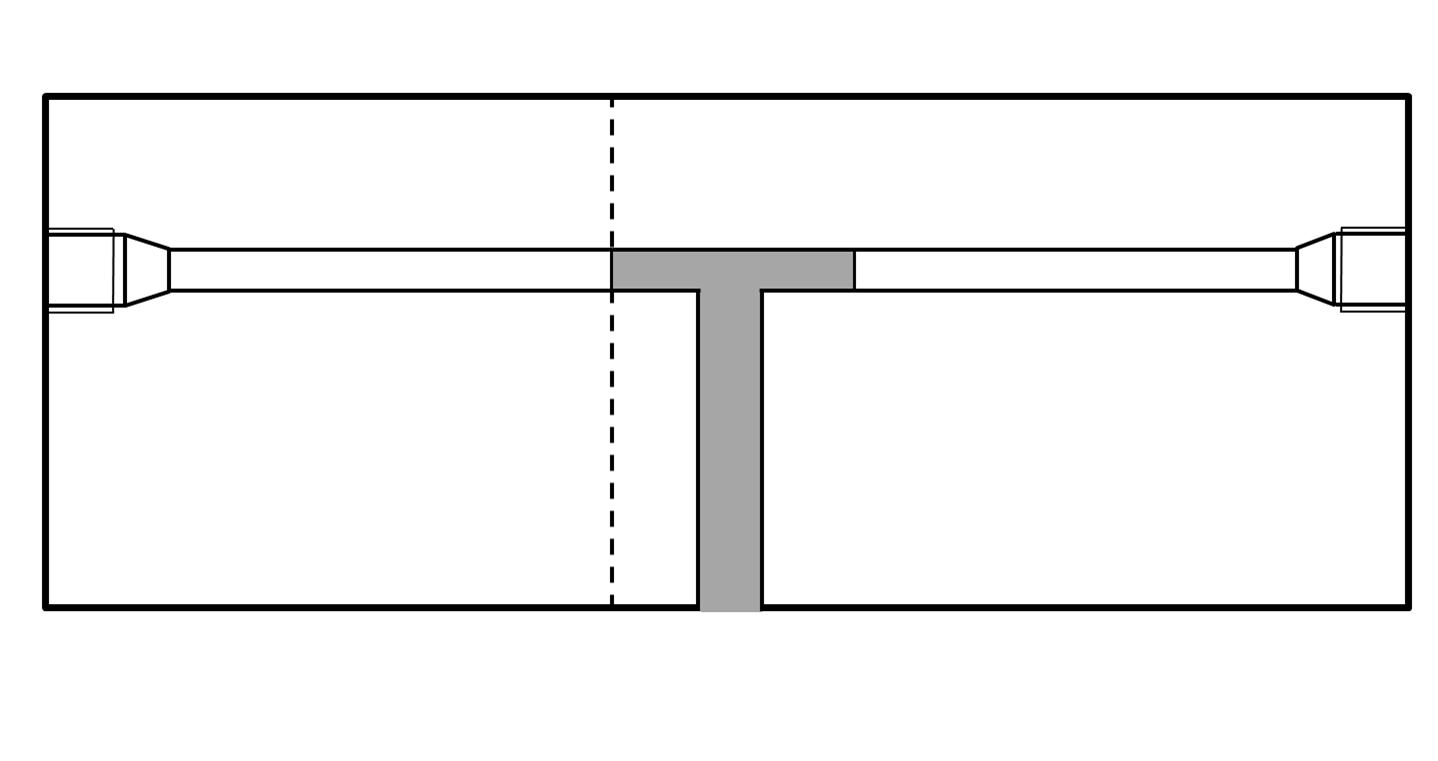}
		\caption{\label{fig:exp_setup}(a) Sketch of the experimental setup. (M), (TC) and (PR) denote the stepping motor, temperature-controlled barrels and the position of the pressure recording, respectively. A zoom of the T-mixer is shown in (b). The grey shaded T-mixer depicts the domain used in the simulations. }

	\end{center}
\end{figure}
Fig.~\ref{fig:exp_setup} shows a sketch of the full experimental setup, which consists of two inlets of square cross section of hydraulic diameter $d=1$\si{mm}, which discharge in a mixing channel of $1$mm height, $2$mm width and $11.5$mm length. The inlets are $18$mm long, whereas in the simulations the inlets are just $3$mm long because we impose fully developed flow as boundary condition. The geometry of the simulations is shown gray shaded in Fig.~\ref{fig:exp_setup}.  Pressure transmitters model A09 (Sensor-Technik Wiedemann GmbH, Germany) with a range up to $0.5$bar relative pressure were used to measure the pressure difference between the fluid in the T-mixer and the environment. To avoid  disturbing the inflow close to the junction, we did not install pressure probes at the locations of the simulated geometry (\ie 3mm before the junction). Instead, the pressure probes were installed before the connections to the inlets of the T-mixer device (PR).  To enable comparison with the simulations, a second device consisting of a square duct of $15$mm length was manufactured (the dashed line in Fig.~\ref{fig:exp_setup}(b) depicts the position its outlet). With this second device, experiments where performed to measured the pressure loss. This was then subtracted from the total loss measured in the experiments with the actual T-mixer, thereby allowing a direct comparison to the simulations. 

Before each experimental run, a baseline pressure signal was recorded for 5 seconds. Then, after the pistons started moving and the flow structures developed, the pressure signal reached a plateau, indicating operation in steady-state conditions. For each experiment, the average pressure of the baseline was subtracted to the average steady-state pressure. Finally, we note that at low Reynolds numbers (up to Re 400) the resulting pressure loss is of the order of the measurement uncertainty. Although deviations are expected, the agreement with numerical simulations is excellent, see Fig.~11(a) in \S~4 (paper).

\section{Villermaux-Dushman Reaction}
\begin{figure}[!htbp]
\begin{center}
\includegraphics[width=0.5\textwidth]{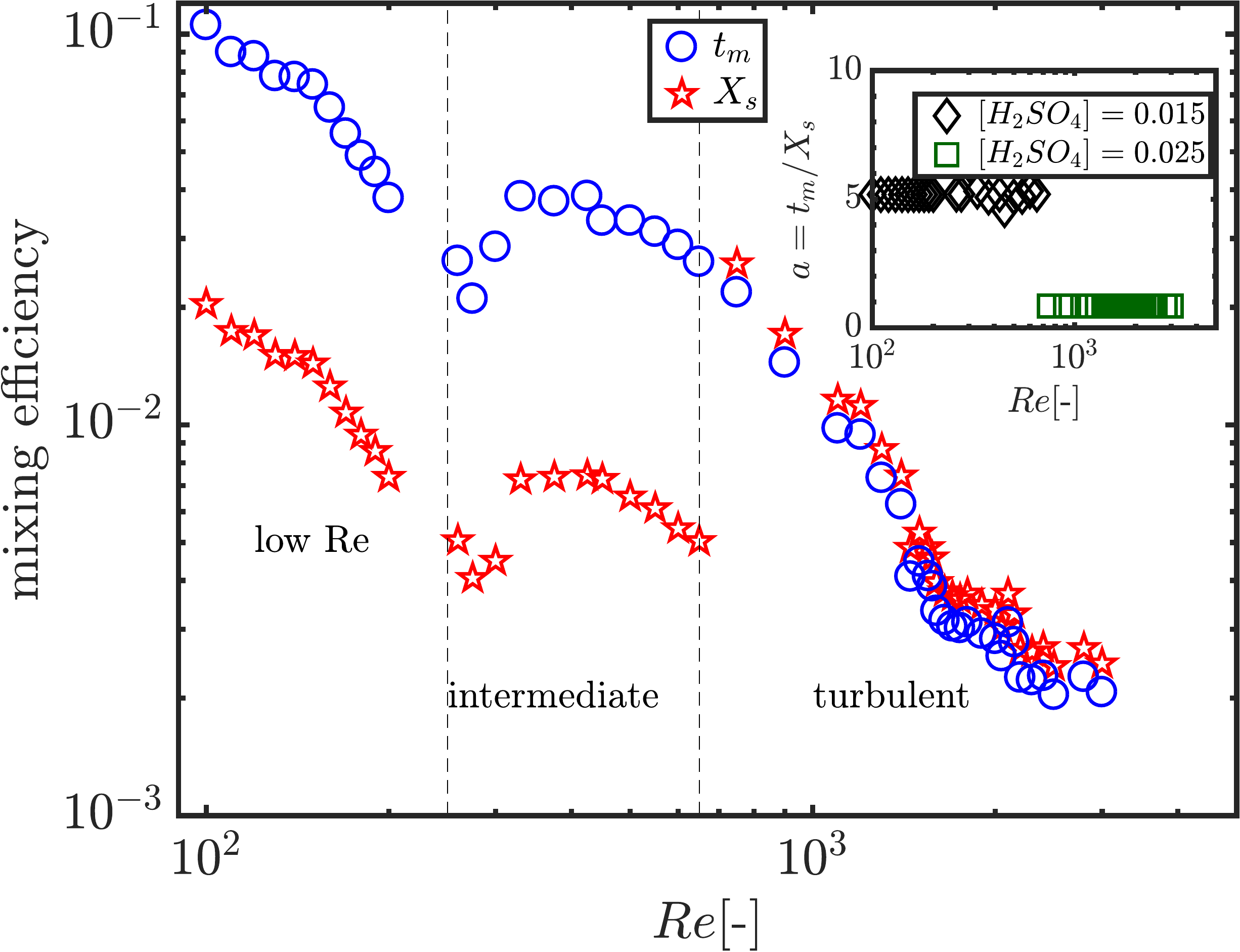}
\caption{\label{fig:segindex}Experimentally obtained mixing efficiencies (segregation index $X_s$[-] and  mixing time $t_m$[s]) as function of the Reynolds number. The inset shows the concentration dependence of $a=t_m/X_s$ as function of Reynolds number $Re$.}
\end{center}
\end{figure}
\subsection{Background}
Chemical reactions, used as molecular probes for mixing, are powerful tools to gain insights into the mixing process. In the literature, several competitive consecutive and parallel chemical reactions have been proposed, where the distribution of quantifiable chemical species depicts the mixing history\cite{paul1971mixing,nabholz1977chemical,yu1993micromixing}. Fournier \etal\ \cite{fournier1996newex} proposed the parallel use of the neutralization of boric acid 
\begin{equation}
H_{2}BO_{3}^{-} + H^{+} \rightleftharpoons H_{3}BO_{3}\label{eq:neut_bor_ac}.
\end{equation}
with the Dushman--reaction\cite{dushman1904rate}
\begin{equation}
5 I^{-} + IO^{-}_{3} + 6 H^{+} \rightleftharpoons 3 I_{2} + 3 H_{2}O\label{eq:dushman},
\end{equation}
namely the Villermaux--Dushman reaction~\cite{commenge2011villermaux}, to quantify the mixing efficiency of mixing devices. In this reaction system, sulfuric acid is added in a stoichiometric deficit. If mixing is nearly perfect, then sulfuric acid is entirely consumed due to the very fast kinetics, whereas if mixing is poor the by-product iodine $I_2$ emerges from the Dushman reaction.
 
While the neutralization of boric acid in Eq.~\ref{eq:neut_bor_ac} is quasi-instantaneous, the kinetics of the Dushman reaction is much slower and in the range of the mixing process \cite{guichardon2000characterisationkin}. The iodine formed in the redox-reaction will further react in a quasi-instantaneous reaction with iodide ions and form an equilibrium with triiodide-ions \cite{guichardon2000characterisationex}, which reads
\begin{equation}
I^{-} + I_{2} \rightleftharpoons I^{-}_{3}.
\label{eq:iodide}
\end{equation}
The concentration of the formed triiodide ions $I^-_3$ was measured here with a Cary 100 Scan UV/Vis spectrometer (Varian Deutschland GmbH) at a wavelength of 353 \si{nm} and quantified by applying the Beer--Lambert law
\begin{equation}
[I_3^-]=\frac{\text{OD}}{\epsilon_{353}\cdot l} \qquad \text{valid if } 0.1<\text{OD}<2.5,
\label{eq:opticaldensity}
\end{equation}
where OD, $\epsilon_{353}$ and $l$ denote the optical density (absorbance), the extinction coefficient of triiodide at wavelength $353$nm and the optical path length within the measurement cell, respectively. Note that the UV/VIS spectrometer was not connected to the mixing setup. Thus the samples collected in a beaker were measured offline. In doing so, at least 1.5ml of the mixture was collected for each experimental trial, while 1ml of this collected amount was needed for optical density measurements itself.

\subsection{Notes on mixing time}

The mixing efficiency is quantified in the Villermaux--Dushman characterization by the segregation index 
\begin{equation}
X_s = \dfrac{Y}{Y_{st}},
\end{equation}
which is the fraction of $H^+$ ions used to form iodine $I_2$, \ie
\begin{equation}
Y = \frac{[I_2]+[I_3^-]}{[H^+]_0} \qquad \text{and} \qquad Y_{st} = \frac{6[IO_3^-]_0}{[IO_3^-]_0+[H_2BO_3^-]_0},
\end{equation}
where the subscript $_0$ denotes the initial concentration. The weakness of this approach lies in the dependence of the segregation index $X_s$ on the initial concentrations. In our experiments, the mixing efficiency changes enormously as $Re$ increases from $100$ to $4000$, and this requires the application of different concentrations sets to stay in the linear regime of the detector of the spectrometer (see Eq.~\ref{eq:opticaldensity}). However, using different concentrations sets results in a discontinuous transition in the segregation index $X_s$ as the Reynolds number increases, see Fig.~\ref{fig:segindex}. Two different concentrations sets are applied in this study (see Tab.~\ref{tab:concentrations}), which are changed at $Re=650$.

Commenge \etal\ \cite{commenge2011villermaux} solved this problem by providing a single empirical a master curve capturing multiple concentration sets. In particular, their master curve converts the measured optical density (effectively the segregation index) into a mixing time 
\begin{equation}
t_{m} = 0.33 (\text{OD})[H^{+}]_0^{-4.55} [I^{-}]_0^{-1.5}[IO_{3}^{-}]_0^{-5.8}[NaOH]_0^{-2}[H_{3}BO_{3}]_0^{-2}.
\label{eq:mixingtime}
\end{equation}
Their approach relies on the interaction of exchange with the mean (IEM) model, which incorporates the influence of the mixing time $t_m$ on the Villermaux-Dushman reaction and the actual reaction system including the kinetics\cite{commenge2011villermaux,falk2010performance}. By varying the mixing time in the model, different segregation indices $X_s$ were obtained and compared to the measured ones. Doing this for different mixers, Reynolds numbers and concentration sets, large datasets were gathered and collapsed together on one master curve by normalizing with the product of reactant concentrations as introduced above in Eq.~\ref{eq:mixingtime}. Two of the concentration sets given in their work enabled the seamless identification of mixing times between $0.1$ and $0.001$s in our work, whereby the actual concentrations are given in Tab.~\ref{tab:concentrations}.

In practice, within one concentration set the mixing time $t_m$ can be obtained by just multiplying a constant factor to the segregation index $X_s$ as 
\begin{equation}
t_m \approx a\cdot X_s,
\end{equation}
where the constant $a$ depends solely on the concentration set as depicted in Fig.~\ref{fig:segindex}. Thus, the herein used mixing time $t_m$ is nothing else than a normalized segregation index.

\begin{table*}
\caption{Base concentrations and acid concentrations used for the Villermaux--Dushman reaction.}
\label{tab:concentrations}
	\begin{center}
		\begin{tabular}{cccc}
		\hline
		\multicolumn{2}{c}{\multirow{2}{*}{Reactants}} & \multicolumn{2}{c}{concentrations \si{mol} \si{l^{-1}}} \\
		\multicolumn{2}{c}{}& $Re=(100,650)$ & $Re=(700,4000)$  \\ \cline{1-4}
				acid&H\textsubscript{2}SO\textsubscript{4} & 0.015 & 0.025 \\\hline
		\multirow{4}{*}{base} &NaI & \multicolumn{2}{c}{0.032}\\
		&KIO\textsubscript{3} & \multicolumn{2}{c}{0.006}\\
		&NaOH & \multicolumn{2}{c}{0.09} \\
		
		&H\textsubscript{3}BO\textsubscript{3} & \multicolumn{2}{c}{0.09}\\\cline{1-4}
		
		\end{tabular}
	\end{center}
\end{table*}
\subsection{Experimental protocols}
Impurities brought into the reaction system can drastically change the kinetics of chemical reactions. This could occur either by unwanted alternative reactions or by merely changing the ionic strength. To detect such problems, out T-mixer was carefully cleaned every day before and after the experimental runs. Each experiment was repeated at least three times and at different days using solutions freshly prepared each day. In all experiments, ultrapure water, produced with a Purelab\textsuperscript{\textregistered} Ultra from Elga LabWater (Veolia Water Solutions \& Technologies, France), was used. For the Villermaux--Dushman characterization, the chemicals used were all of high purity, either analytical grade (p.a.\ or Reag.\ PhEur) or of 99.8\% purity. Sulfuric acid 96\% (p.a.), potassium iodate (p.a.) (Carl Roth GmbH, Germany), sodium iodite (Reag. Ph Eur) (Merk KgaA, Germany), sodium hydroxide (reagent grade)(Sigma-Aldrich Chemie GmbH, Germany) and boric acid (99,8\%) (Alfa Aesar, Germany) were used. The solutions were prepared as described by Guichardon \etal\ \cite{guichardon2000characterisationex} to prevent thermodynamic triiodide formation. Additionally, before the solutions were prepared, the ultrapure water was stripped with nitrogen gas to avoid the oxidation of iodide to iodine \cite{guichardon2000characterisationkin}. The employed chemicals do not influence the material parameter of the aqueous solution, \eg viscosity and density, and thus enable to compare the results qualitatively between simulation and experiment. 

\noindent{\bf Additional References}
\bibliographystyle{plain}
\bibliography{biblio_Holger}